\documentclass[10pt,iop]{emulateapj}
\usepackage{amsmath}
\pdfoutput=1

\begin{document}

\title{The Baryon Acoustic Oscillation Broadband and Broad-beam Array: Design Overview and Sensitivity Forecasts}
\author{Jonathan C. Pober\altaffilmark{1}, 
Aaron R. Parsons\altaffilmark{1}, 
David R. DeBoer\altaffilmark{2},
Patrick McDonald\altaffilmark{3},
Matthew McQuinn\altaffilmark{1,4},
James E. Aguirre\altaffilmark{5},
Zaki Ali\altaffilmark{1},
Richard F. Bradley\altaffilmark{6,7},
Tzu-Ching Chang\altaffilmark{8},
Miguel F. Morales\altaffilmark{9}}

\altaffiltext{1}{Astronomy Dept., U. of California, Berkeley, CA}
\altaffiltext{2}{Radio Astronomy Laboratory, U. of California, Berkeley, CA}
\altaffiltext{3}{Physics Div., Lawrence Berkeley National Laboratory, Berkeley, CA}
\altaffiltext{4}{Einstein Fellow}
\altaffiltext{5}{Dept. of Physics and Astronomy, U. of Pennsylvania, Philadelphia, PA}
\altaffiltext{6}{Astronomy Dept. and Dept. of Electrical and Computer Engineering, U. of Virginia, Charlottesville, VA}
\altaffiltext{7}{National Radio Astronomy Observatory, Charlottesville, VA}
\altaffiltext{8}{Institute of Astronomy and Astrophysics, Academia Sinica, Taipei, Taiwan}
\altaffiltext{9}{Dept. of Physics, U. of Washington, Seattle, WA}

\begin{abstract}
This work describes a new instrument optimized for a detection of
the neutral hydrogen 21cm power spectrum between redshifts of $0.5-1.5$: the 
Baryon Acoustic Oscillation Broadband and Broad-beam (BAOBAB) Array.  
BAOBAB will build on the efforts of a first generation of 21cm experiments
which are targeting a detection of the signal from the Epoch of Reionization at
$z\sim10$.  
At $z\sim1$, the emission from neutral hydrogen 
in self-shielded overdense halos also presents an accessible signal, 
since the dominant, synchrotron foreground emission is considerably fainter
than at redshift 10.  The principle science driver for these observations are
Baryon Acoustic Oscillations in the matter power spectrum
which have the potential to act as a standard ruler and constrain the nature of dark energy.
BAOBAB will fully correlate
dual-polarization antenna tiles over the 600--900MHz band with a frequency resolution of 
300 kHz and a system temperature of 50K.  The number of antennas will grow in 
staged deployments, and reconfigurations of the array
will allow for both traditional imaging and high power spectrum sensitivity operations.  We present
calculations of the power spectrum sensitivity for various array sizes, 
with a 35-element array measuring the cosmic neutral hydrogen 
fraction as a function of redshift, 
and a 132-element system detecting the BAO features in the power spectrum,
yielding a $1.8\%$ error on the $z\sim1$ distance scale, and, in turn, significant
improvements to constraints on the dark energy equation of state over an
unprecedented range of redshifts from $\sim~0.5-1.5$.
\end{abstract}

\keywords{instrumentation: interferometers --- cosmological parameters --- distance scale --- techniques: interferometric --- large-scale structure of the universe}

\section{Introduction}

The Baryon Acoustic Oscillation (BAO) features in the large-scale
matter distribution have recently drawn attention as a standard ruler
by which the geometry of the universe can be directly measured 
\citep{eisenstein_et_al_1998,eisenstein_et_al_1999}.
These features in the cosmic microwave background (CMB)
power spectrum and the matter power spectrum today are
imprints from the acoustic oscillations in the primordial photon-baryon plasma
that recombined at $z\approx1100$. The features in the power spectrum
appear at multiples of 
the sound horizon scale at recombination, making them effective 
standard rulers.  Measuring the BAO wiggles at several
redshifts yields geometric measurements of the universe --- the Hubble
parameter, $H(z)$, and the angular diameter distance, $d_A(z)$ --- that
constrain properties of the dark energy that dominates the cosmic energy
content at $z=0$ and is the current leading theory for the accelerated expansion
of the universe.  Since the first detection of the BAO signal
\citep{eisenstein_et_al_2005}, several experiments have been undertaken
to use these features for precision cosmology, including the Sloan
Digital Sky Survey III Baryon Oscillation Spectroscopic Survey
(SDSS-III BOSS;
\citealt{schlegel_et_al_2009})\footnote{http://www.sdss3.org/surveys/boss.php/},
WiggleZ
\citep{drinkwater_et_al_2010}\footnote{http://wigglez.swin.edu.au/site/},
and the Hobby-Eberly Telescope Dark Energy Experiment
(HETDEX; \citealt{adams_et_al_2011})\footnote{http://hetdex.org/},
as well as a number of planned future experiments, such as the
Subaru Prime Focus Spectrograph (PFS; \citealt{ellis_et_al_2012}),
Euclid \citep{amendola_et_al_2012}, BigBOSS \citep{schlegel_et_al_2011},
and the Wide-Field Infrared Survey Telescope (WFIRST)\footnote{http://wfirst.gsfc.nasa.gov/}.
All of these experiments target individual galaxies with spectroscopic
observations.

Rather than targeting individual objects, a 21cm intensity mapping
experiment can detect fluctuations in neutral hydrogen emission on large scales
\citep{chang_et_al_2008,wyithe_et_al_2008,morales_and_wyithe_2010,pritchard_and_loeb_2012},
with two dimensions corresponding to angles
on the sky, and the third line-of-sight dimension arising from the differential 
redshifting of 21cm line emission as a function of distance.
After reionization, the power spectrum of 21cm fluctuations is expected to be a biased tracer of the matter power-spectrum,
since the remaining neutral hydrogen resides in
high-density, self-shielded regions such as in galaxies and other collapsed halos
\citep{barkana_and_loeb_2007,madau_et_al_1997}.  As a result, 
21cm intensity mapping experiments
present a promising complement to spectroscopic galaxy surveys
for BAO science.  Several 21cm intensity mapping experiments have been proposed,
including the prototype Cylindrical Radio Telescope (CRT; formerly HSHS, 
\citealt{peterson_et_al_2006}; \citealt{seo_et_al_2010})\footnote{\url{http://cmb.physics.wisc.edu/people/lewis/webpage/}},
the Canadian Hydrogen Intensity Mapping Experiment (CHIME)\footnote{http://www.mcgillcosmology.ca/chime},
BAORadio \citep{ansari_et_al_2012a,ansari_et_al_2012b}, 
the BAO from Integrated Neutral Gas Observations experiment (BINGO; \citealt{battye_et_al_2012},
and an ongoing experiment with the Green Bank
Telescope (GBT; \citealt{chang_et_al_2010}).  

The flexibility in angular and spectral responses of radio interferometers,
which measure the power spectrum both parallel and perpendicular to
the line of sight \citep{morales_2005},
gives 21cm BAO the ability to survey larger cosmological volumes and operate over
a wider range of redshifts than current spectroscopic galaxy redshift surveys.
As a result, the 21cm BAO signal has the potential for probing expansion throughout
and beyond
the critical epoch when dark energy comes to dominate the energy density of the
universe.
Furthermore, a 21cm intensity mapping experiment can probe redshifts $z > 0.5$ 
with roughly uniform sensitivity,
without complications arising from sky emission lines in the optical/near-infrared.
The 21cm signal can be used to constrain the location of the
BAO peaks as a function of redshift, and thereby measure
the magnitude and time-evolution of dark energy.

21cm BAO experiments can draw on the considerable investments in
low-frequency radio astronomy developed in the past decade for studies of
the Epoch of Reionization (EoR).
In this paper, we present the Baryon Acoustic Oscillation Broadband and Broad-beam (BAOBAB)
array, a new experiment, building on the legacy of the Precision Array for Probing
the Epoch of Reionization (PAPER; \citealt{parsons_et_al_2010})\footnote{http://eor.berkeley.edu/} and the Murchison Widefield 
Array (MWA; \citealt{lonsdale_et_al_2009})\footnote{http://www.mwatelescope.org/}, 
for measuring the 21cm HI power spectrum at a redshift of $z\sim1$.
This paper is structured as follows: in \S\ref{sec:baobab}, 
we present a system architecture for the BAOBAB instrument.
In \S\ref{sec:predictions}, we forecast the sensitivity and cosmological
constraints that will be achieved by BAOBAB.
We consider several possible challenges  and extensions for this approach in \S\ref{sec:discussion}, 
and conclude in \S\ref{sec:conclusions}.  Throughout this work 
we assume the WMAP7 best fit $\Lambda$CDM cosmological model:
$h = 0.7$, $\Omega_M = 0.27$, $\Omega_b = 0.046$, $\Omega_{DE} = 0.73$, and $\sigma_8 = 0.8$ 
\citep{larson_et_al_2011}.

\section{The BAO Broadband and Broad-beam Array}
\label{sec:baobab}

The past decade has seen significant progress in the design, construction,
and calibration of low-frequency interferometric arrays toward the goal
of detecting the highly-redshifted 21cm signal from the Epoch of Reionization.
The technologies used in BAOBAB inherit from two EoR experiments ---
PAPER and the MWA --- but with several significant modifications to optimize the instrument
for BAO science.  The entire signal chain will be re-tuned to operate between 
600--900~MHz.  These frequencies corresponds to redshifted 21cm emission between $z = 0.58-1.37$,
a band chosen for several reasons.  First, these moderate redshifts complement the undergoing lower
redshift galaxy surveys like BOSS by probing what is currently a relatively unexplored volume of the universe.
Secondly, at these frequencies, commercially available amplifiers and cables provide suitable
low-noise performance that would not be obtainable at higher frequencies (lower redshifts).  
Furthermore, this band also avoids the bright sky noise and ionospheric effects that complicate
lower frequency (higher redshift) observations. 

\begin{figure}\centering
\includegraphics[width=3in]{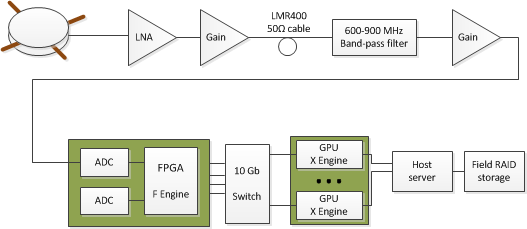}
\caption{
System diagram of the BAOBAB interferometer.  Dual-polarization antenna signals
at -103dBm enter an uncooled low-noise amplifier (LNA) with +12.5dB
gain and a noise figure of 0.4dB (30K).  Second-stage amplifiers add
+36dB gain before transmission through 30 meters of LMR400 50$\Omega$ cable to
a central enclosure.  The signal is bandpass filtered (600--900MHz) and
amplified +40dB to the optimal -22dBm input level for the ADCs.
Each antenna signal is digitized and channelized in ROACH F-engines, reordered
in transmission through 
a 10Gb Ethernet switch, and sent to GPU GTX580 X-engines
for cross-correlation. Raw visibility data are passed to a host
computer and written to a RAID storage unit in the MIRIAD UV file format
for post-processing.
}
\label{fig:signal_flow}
\end{figure}

BAOBAB will be a non-phase-tracking, broadband array of beam-formed tiles.
By lowering the field-of-view of each element,  
the use of beamformed tiles like the MWA will significantly increase BAOBAB's power spectrum sensitivity
over that of an array of the equivalent number of single dipoles, without increasing the correlator demands.
Each tile will consist of 4 scaled versions of the PAPER sleeved-dipole design 
and groundscreens, electronically beamformed to point to zenith.
Two linear polarization signals
from each tile enter a digital signal processor that computes both auto-
and cross-correlation products and outputs the results locally to disk.
A block diagram is given in Fig. \ref{fig:signal_flow}; 
the key properties of the BAOBAB system are listed in Table \ref{tab:baobab}
and are described in more detail in the remainder of this section.

\begin{table}[ht]\centering
\caption{Proposed BAOBAB Array}
\begin{tabular}{l|l}
\hline\hline
Operating Bandwidth & 600--900 MHz \\
Number of Tiles & 32--128 \\
Collecting Area per Element & 2.6 $\rm{m}^2$ \\
Gain per Element & 18 dBi \\
Field-of-View & 0.045 str \\
Receiver Noise Temperature & 40 K \\
System Temperature & 50 K\\
Maximum Imaging Baseline & 60 m\\
Redundant Baseline Scale & 1.6 m\\
$k_{\rm min}$, $k_{\rm max}$ & 0.01, 2.5~$h{\rm Mpc}^{-1}$\\ 
Array Configuration & Reconfigurable (see Fig. \ref{fig:configuration})\\
Frequency Resolution & 300 kHz\\
Snapshot Integration Time & 10 s\\
\hline
\end{tabular}
\label{tab:baobab}
\end{table}

\subsection{Siting}

\begin{figure}\centering
    \includegraphics[height=2.0in]{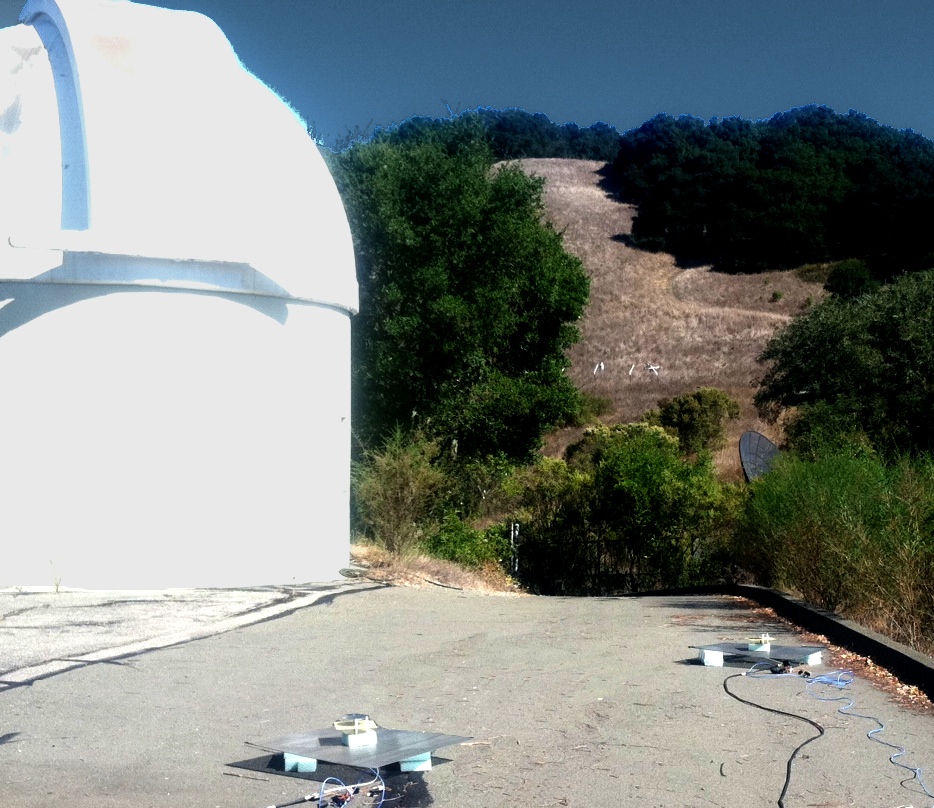}
    \includegraphics[width=3.0in]{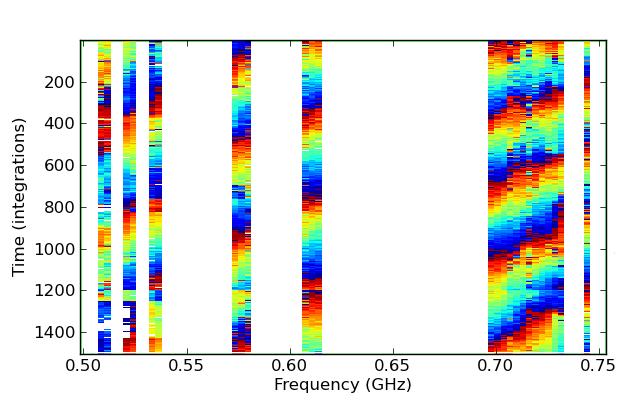}
    \caption{\emph{Top}: Leuschner Observatory, with a prototype 2-element BAOBAB
interferometer deployed.  This system was developed and deployed by students as
part of a {\it Fundamentals of Radio Astronomy} class at UC Berkeley.  \emph{Bottom}:
solar fringes measured with the BAOBAB-2 prototype at Leuschner.
}
\label{fig:leuschner}
\end{figure}

Since radio-frequency interference (RFI) is prevalent at these frequencies,
BAOBAB will need to be located at a radio quiet site.  The bottom
panel of Figure \ref{fig:leuschner} shows preliminary measurements made by a
prototype 2-antenna BAOBAB interferometer deployed at the Leuschner Observatory
near Berkeley, CA.  
At this site, only 40 MHz of a 400--800 MHz operating band
show solar fringes uncorrupted by RFI, demonstrating the need for the primary
BAOBAB deployment to be located 
at a quieter site, such as the NRAO site near Green Bank, WV.  
Next-generation activities may
take place at the Square Kilometer Array South Africa
(SKA-SA) reserve in the Karoo desert.  This site is currently
occupied by the PAPER and MeerKAT arrays, and has been shown to be a pristine RFI
environment \citep{jacobs_et_al_2011}.

\subsection{Analog System}

With the drastic reduction in sky noise relative to EoR frequencies, BAOBAB's system temperature
will be dominated by the analog electronics.
These components must therefore be optimized to reduce
receiver noise
while maintaining the
smooth spatial and spectral responses that are a hallmark of the PAPER design
and a key component of the delay spectrum foreground isolation approach presented in 
\citet{parsons_et_al_2012b} (hereafter P12b) and discussed in \S\ref{sec:dspec}.
The analog system will include the collecting element (consisting of 4 antennas and 
reflectors), low-noise amplifier, coaxial cable, and receiver.

\begin{figure}\centering
\includegraphics[width=2.0in]{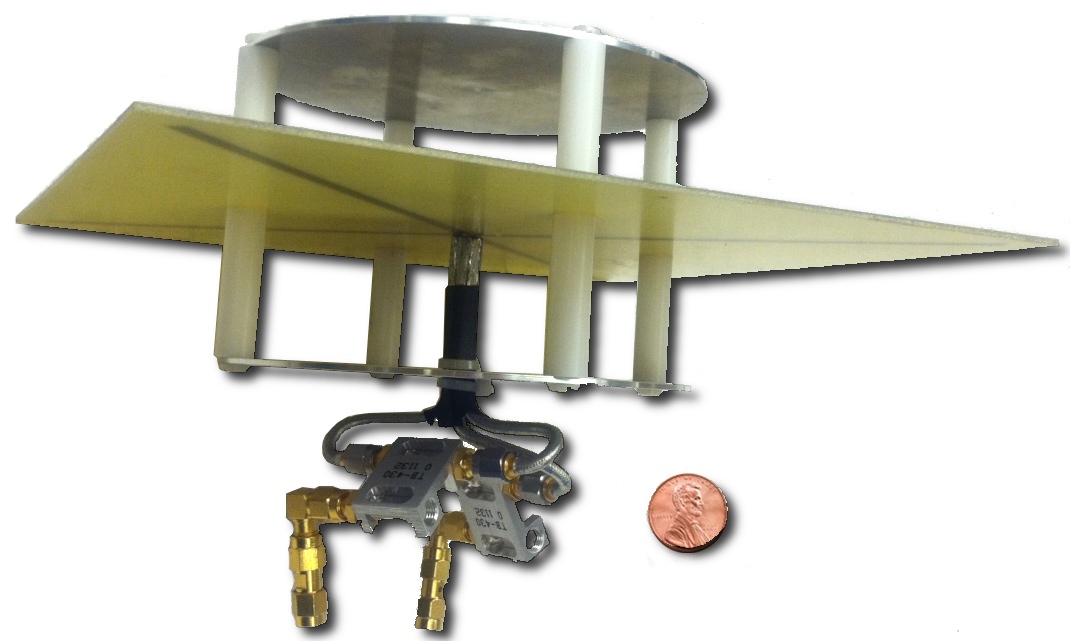}
\includegraphics[height=1.75in]{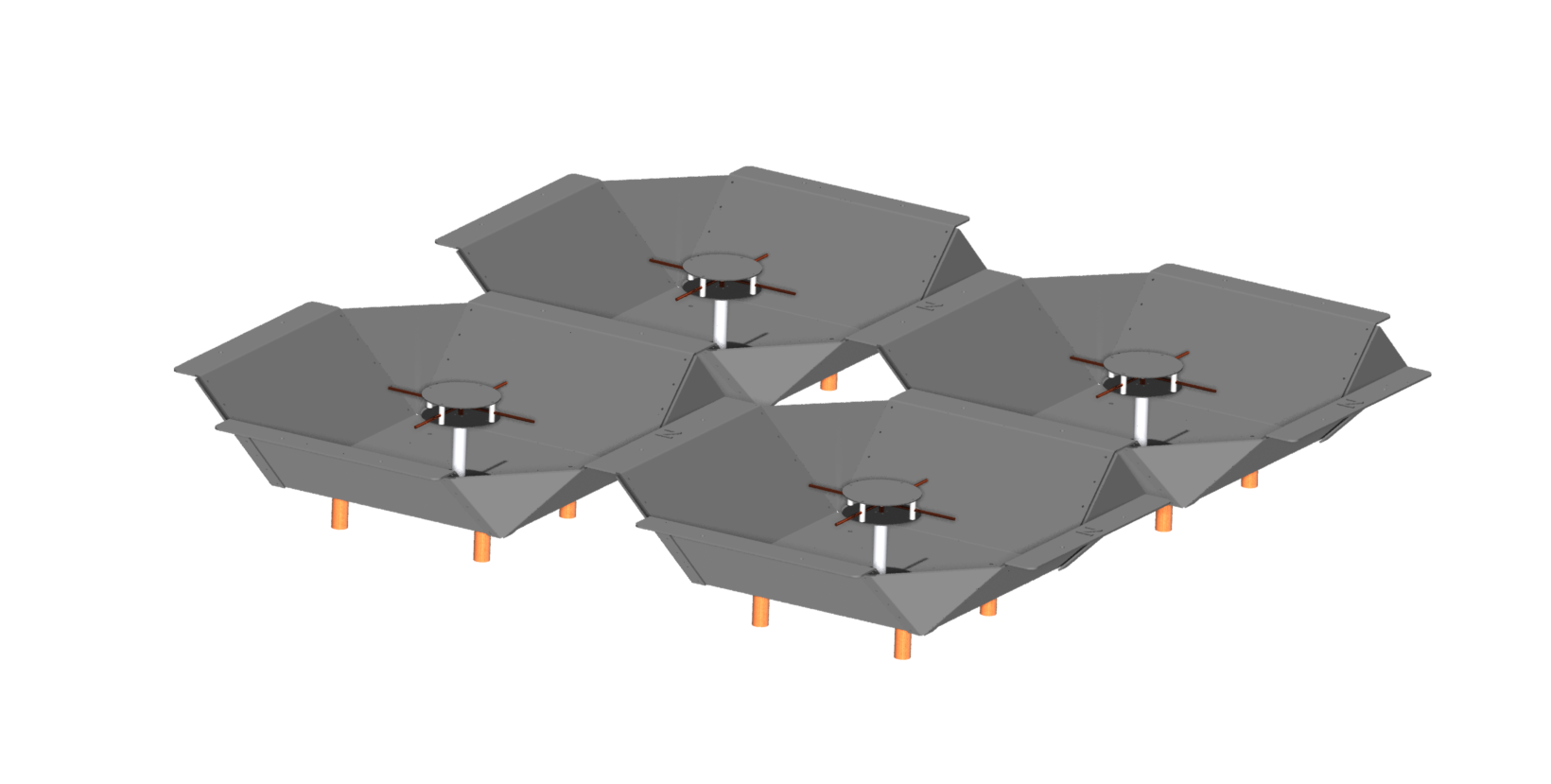}
\caption{
\emph{Top}: A prototype BAOBAB dipole antenna, designed as a 1/5 scale model of a PAPER dipole.
\emph{Bottom}: BAOBAB tile design with 4 dipoles and individual ground-screens.
}
\label{fig:element}
\end{figure}

The BAOBAB element will begin with a 1/5-scale PAPER antenna
\citep{parsons_et_al_2010}, as shown in Figure \ref{fig:element}.  
This design
is a dual-polarized version of the sleeved dipole design that uses a
twin-resonance structure consisting of a pair of crossed dipoles located
between a pair of thin aluminum disks.
The element's
reliability has been demonstrated in PAPER arrays over the past several years.
A trough reflector under each dipole will be used to increase the directivity
toward zenith.  The electromagnetic behavior of the element was modeled extensively for
PAPER using CST Microwave Studio,
and shown to perform as desired through calibration with celestial sources in
\cite{pober_et_al_2012}.  The geometrically re-tuned prototype shown in the
top panel of Figure \ref{fig:element}
will be optimized to operate efficiently over the 600--900 MHz band.

Rather than deploy single elements like PAPER, BAOBAB will use a $2\times2$ tile
of dipoles and ground-screens, as shown in Figure \ref{fig:element}.  A fixed zenith beamformer 
will be used to combine the signals, increasing the gain by 6 dB and reducing the 
field-of-view by a factor of four.  Both analog and digital beamformers are being investigated.  
A key issue is the mutual coupling, which should be reduced by the additional groundscreens between 
dipoles.  The net effect is that for a fixed correlator size, the power-spectrum
sensitivity is increased by a factor of four (see \S\ref{sec:sensitivity}).

\begin{figure}\centering
\includegraphics[height=2in]{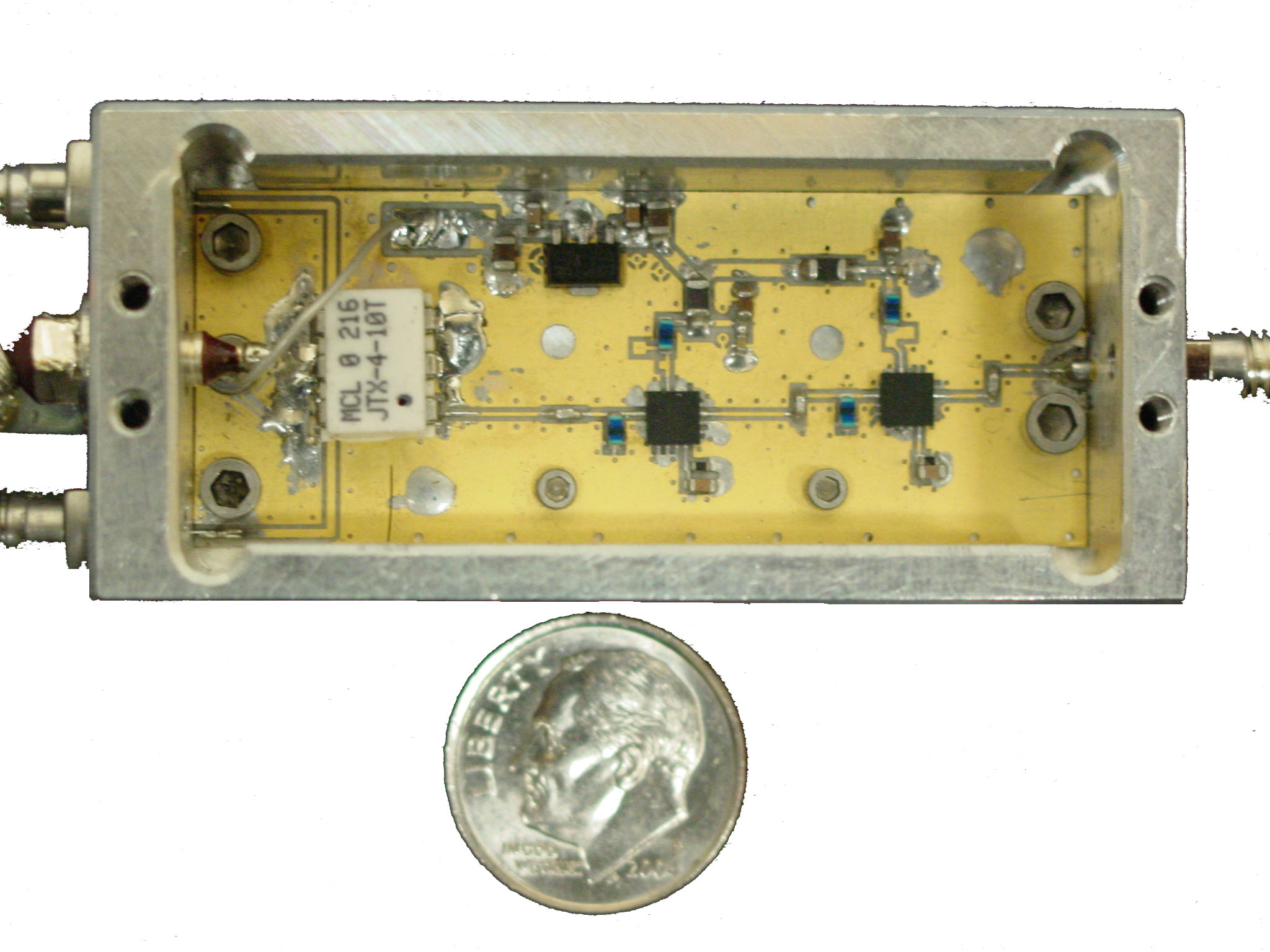}
\caption{The prototype balun and Hittite HMC617LP3 LNA for BAOBAB.  The amplifier adds +30dB of gain
with a quoted noise figure of 0.5 dB.}
\label{fig:balun_rx}
\end{figure}

The amplifier designed for PAPER has a measured noise
temperature of 110 K with 30 dB of gain across the 120-170 MHz band \citep{parsons_et_al_2010}.
For application to BAO at $z\sim1$, we will modify this amplifier design to
operate from 600--900 MHz.  Besides re-tuning the filter and amplifier
circuits, however, one of the major activities in this modification will be
to reduce the noise temperature of the front-end amplifier in order to 
obtain a target system temperature of 50~K.
This change reflects one of the key differences between the BAO and
EoR foregrounds. System noise in the EoR band is dominated by $\sim\!\!300$ K sky noise
from galactic synchrotron emission.  In the BAO band, the sky temperature is reduced
to $\sim\!\!10$ K, making the front-end amplifier the leading source of noise.
Uncooled commercial UHF-band amplifier transistors based on GasFET or HEMT technology can
reliably achieve noise figures of 0.4 dB, corresponding to a receiver temperature of 30K.
A prototype BAOBAB balun/amplifier using a Hittite HMC617LP3 LNA with a quoted noise 
figure of 0.5 dB is shown in Figure \ref{fig:balun_rx}; tests are underway to determine
the noise temperature of the complete system.

\subsection{Digital System}

\begin{figure}\centering
\includegraphics[height=2.3in]{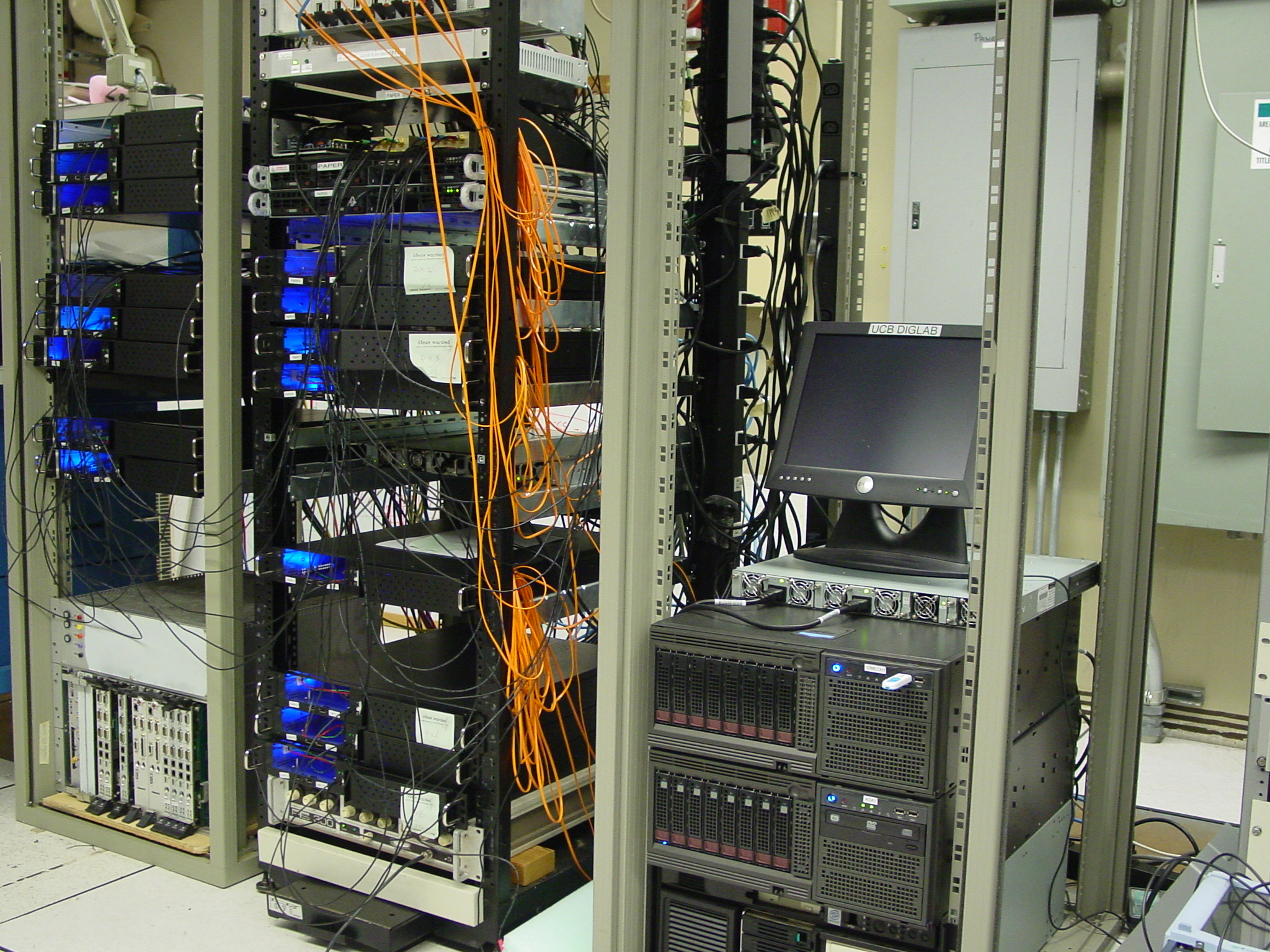}
\caption{
PAPER's 128-input correlator (shown) follows the packetized frequency--cross-multiply (FX)
architecture developed by the Center for Astronomy Signal Processing and
Electronics Research (CASPER).  Shown are 16 ROACH F-engines (left)
and 2 dual-GPU box X-engines (right).  The first-generation BAOBAB correlator will
modify the 64-input, 100-MHz PAPER correlator to become a 32-input,
300-MHz correlator.  It will employ eight ROACH boards for spectral processing
and four dual-GPU boxes for cross-multiplication.  A 10-GbE switch is used to
route data between boards.
} \label{fig:correlator}
\end{figure}

The BAOBAB correlator will follow the 
scalable correlator design used by PAPER and
other members of the international Collaboration for Astronomy Signal
Processing and Electronics Research (CASPER)\footnote{https://casper.berkeley.edu}, a real-time digital correlator
based on Field-Programmable Gate Array (FPGA) processors and graphics processing units (GPUs)
\citep{parsons_et_al_2008,clark_et_al_2011}.
The correlator architecture we employ uses modular
signal processing electronics and packetized communication protocols to
build correlators that are flexible in the number of antennas correlated and
the bandwidth correlated by each antenna.
A photograph of a 128-input FPGA/GPU correlator is shown in Fig. \ref{fig:correlator}.

The generic FX correlator architecture we implement consists of
modules responsible for digitizing and
channelizing each set of antenna inputs (F-Engines), followed by a set
of signal processing modules responsible for cross-multiplying all
antennas and polarizations for a single frequency (X-Engines) and
accumulating the results. 
Unique to this architecture, 
signal processing engines transmit packetized data through
commercially available 10 Gb Ethernet switches
that are responsible for routing data between boards.  This
architecture, along with analog-to-digital converters, modular
FPGA-based signal processing engines, and a software environment for
programming, debugging, and running them, were developed in
collaboration with CASPER at the University of California, Berkeley \citep{parsons_et_al_2008}.  
The flexibility and modularity of this correlator design
shortens development time, and in this case, allows an existing 64-input, 100-MHz
PAPER correlator with 8 ROACH-boards and 4 dual-GPU boxes
to be straight-forwardly modified to become a 32-input, 300-MHz BAOBAB correlator
using the same boards and signal processing libraries.  A forthcoming publication
on this correlator will be presented in Ali et al. (2013).
 
\subsection{Configuration}
 
BAOBAB will employ small antennas and above-ground cabling with relatively inexpensive 
LMR400 50-Ohm coaxial cables; these cables will not be buried, allowing BAOBAB
to easily change between different array configurations by moving antenna elements.
Following the principles outlined in
(\citealt{parsons_et_al_2012a}, hereafter P12a), BAOBAB will employ a minimum-redundancy imaging 
configurations for characterizing foregrounds with minimal sidelobes
and maximum-redundancy configurations to repeatedly
sample the same locations in the $uv$-plane, substantially improving sensitivity
to the three-dimensional power spectrum of 21cm emission at $z\sim1$.
Although future experiments may target a range of angular scales to map 21cm emission
in the plane of the sky, by focusing on a limited number of Fourier modes, 
these
maximum-redundancy configurations can improve sensitivity to the
power spectrum by an order of magnitude or more in mK$^2$, relative to an
equivalent observation with a minimum-redundancy configuration.

\begin{figure}\centering
\includegraphics[width=1.75in]{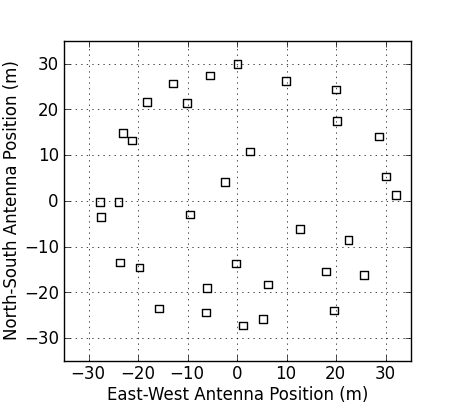}\includegraphics[width=1.75in]{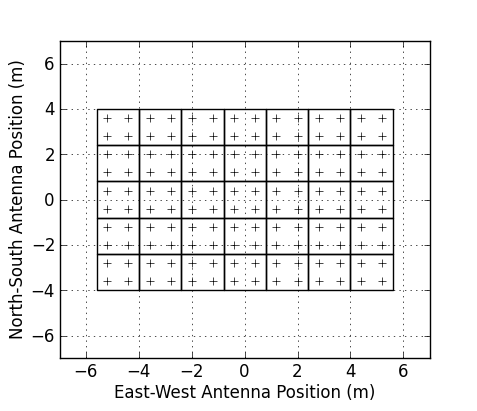}
\caption{BAOBAB array configurations, plotted in meters east-west 
(horizontal axis) and north-south (vertical axis).  BAOBAB will use above-ground
cabling to allow antennas to be moved into
a minimum-redundancy configuration (left) for imaging foregrounds, or
a maximum-redundancy configuration (right) for enhanced power-spectrum sensitivity.
Each square represents one tile, and each ``+" one dipole.
}
\label{fig:configuration}
\end{figure}

However, as will be discussed further in \S\ref{sec:thermal_noise}, the mapping
of
baseline length to a transverse $k$-mode on the sky is significantly larger for
BAOBAB
than for PAPER.  In order to probe the relatively large-scale BAO features, 
then, BAOBAB will
use the most compact configurations possible for its power spectrum 
measurements.
Such an array configuration for a 35-element system, as well as that of a 
32-element 
imaging configuration, 
are shown in Figure \ref{fig:configuration}.  
The tiles are spaced 1.6m apart, effectively touching 
end-to-end.  Investigations of cross-talk and mutual coupling will take place during an early
prototype of the system; it may be the case that a phase-switch or additional shielding between
tiles will be necessary to accommodate the short baselines required by BAO science.

With a modular CASPER correlator increasingly able to process larger numbers of antenna inputs, 
BAOBAB naturally lends itself to a staged approach.  Early $\leq 16$ tile-element prototypes 
will characterize system performance, while a subsequent
$\sim32$-element array will study foreground emission and 
constrain the neutral hydrogen fraction as a function of redshift 
with a measurement of the 21cm power
spectrum (\S\ref{sec:baobab32}).  A $\sim128$-tile version of BAOBAB 
will measure BAO features and provide substantial improvements over our current
constraints on the equation of state and time evolution of dark energy (\S\ref{sec:baobab128}). 

\section{Predicted Cosmological Constraints from BAOBAB}
\label{sec:predictions}

In this section we present predictions for forthcoming cosmological constraints for
several iterations of the BAOBAB instrument.  We begin by reviewing the predicted
signal strength for the cosmological 21cm power spectrum in \S\ref{sec:ps21}.
In \S\ref{sec:sensitivity}, we adapt the power spectrum sensitivity calculations of 
P12a for an array operating at $z\sim 1$, including the effects of sample variance
and shot noise.  In \ref{sec:dspec}, we briefly review the delay spectrum
foreground removal procedure of P12b.
While a detailed study of foregrounds is beyond the scope of this paper,
it is worthwhile to discuss the implications of the technique on which
Fourier modes of the 21cm power spectrum will be accessible.
We conclude the section by presenting forecasts for a 35- and 132-element BAOBAB
system in \S\ref{sec:baobab32} and \S\ref{sec:baobab128}
respectively, including Fisher matrix predictions for constraints on the dark energy 
equation of state in the latter section.  In the discussion of \S\ref{sec:discussion},
we explore possible directions for improvement with larger BAOBAB arrays.

\subsection{The 21cm Power Spectrum}
\label{sec:ps21}

As with galaxy redshift surveys, a 3D map of the neutral hydrogen in the universe
will serve as a tracer of the underlying dark matter power spectrum.  The brightness 
of the observable radio 21cm signal will depend on the cosmological neutral hydrogen
fraction, as well as the bias of hydrogen containing halos with respect to the dark matter
\citep{barkana_and_loeb_2007,madau_et_al_1997,ansari_et_al_2012b}:
\begin{equation}
    \label{eq:pred_sig_b}
    P_{T_{21}}(k,z) = \left[\tilde T_{21}(z)\right]^2 b^2 P(k,z),
\end{equation}
\begin{align}
    \label{eq:pred_sig_fhi}
    \tilde T_{21}(z) \simeq 0.084 {\rm mK} \frac{(1+z)^2 h}{\sqrt{\Omega_m(1+z)^3+\Omega_\Lambda}} \frac{\Omega_B}{0.044}\frac{f_{\rm{HI}}(z)}{0.01},
\end{align}
where $\tilde T_{21}(z)$ is the mean 21cm brightness temperature at
redshift $z$; $P(k,z)$ is the linear matter power spectrum; $b$ is the
bias factor of HI containing halos with respect to the dark matter;
$f_{\mathrm{HI}}(z)$ is the mass fraction of neutral hydrogen with
respect to the overall cosmological baryon content (i.e., $\Omega_{\rm{HI}} = f\Omega_b$); $\Omega_\Lambda$
is the cosmological constant, and $\Omega_m$ and $\Omega_B$ are the
matter and baryon density in units of the critical density,
respectively.

\begin{figure}\centering
\includegraphics[width=2.5in]{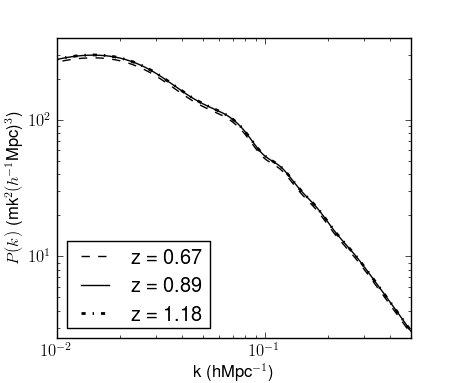}
\includegraphics[width=2.5in]{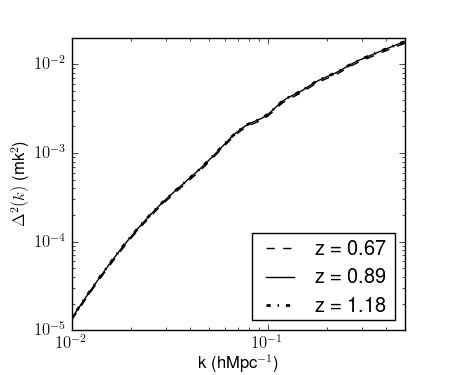}
\caption{
\emph{Top}: The predicted real space (i.e., $\mu=0$) power spectrum of 21cm 
emission in our fiducial model
at three redshifts:
0.67, 0.89 and 1.18 (corresponding to frequencies of 850, 750 and 650 MHz).  
The $z = 0.89$ and $z = 1.18$ power spectra overlap, and, effectively, the 
signal is at the same strength for all three  
redshifts.  Matter power
spectrum predictions come from CAMB.  \emph{Bottom}: The equivalent 
dimensionless power spectra,
$\Delta^2(k) = \frac{k^3}{2\pi^2}P(k)$.
} \label{fig:21cmps}
\end{figure}

We plot the predicted 21cm 
brightness temperature power spectrum, $P(k)$, for our fiducial model 
at redshifts of 0.67, 0.89 and 1.18 (corresponding to frequencies of 850,
750, and 650 MHz, respectively) in Figure \ref{fig:21cmps}.  
Predictions for the matter power spectrum
come from CAMB \citep{lewis_et_al_2000}\footnote{http://camb.info/}.  
We also plot the dimensionless power
spectrum, $\Delta^2(k) = \frac{k^3}{2\pi^2}P(k)$.  For the remainder of this paper
will we primarily express our results in terms of $\Delta_{\rm{T_{21}}}^2(k)$, due to its more
intuitive units (mK$^2$) and simple physical interpretation as the variance
per logarithmic $k$ bin.

In practice, the 21cm power spectrum is measured in
redshift space, which can be, at linear order, related to the real-space
$P_{T_{21}}(k,z)$ as \citep{kaiser_1987}:
\begin{equation}
\label{eq:rsd}
\tilde{P}_{T_{21}}(k,z)=(1 + \beta \mu^2)^2 P_{T_{21}}(k,z)
\end{equation}
where $\mu = \hat{k} \cdot \hat{z}$ is the wavevector $\hat{k}$
projected along the line-of-sight $\hat{z}$, and $\beta \equiv f(\Omega)/b$ from
linear theory where $f(\Omega) \approx \Omega_m(z)^{\gamma}$ is the
dimensionless linear growth rate and $\gamma=0.557$ for $\Lambda$CDM
cosmologies.  To incorporate the effects of redshift-space distortions in 1D plots of the power
spectrum, we reduce our thermal noise error bars in $k$-space by the factor of $(1+\beta\mu^2)^2$;
however, our cosmological analyses keep full 2D information.
We do not attempt to constrain cosmological parameters like $\beta$ by measuring the power
spectrum as a function of $\mu$.  
We assume fiducial values of $f_{\rm{HI}}$ =
0.015 and $b$ = 0.75, chosen to agree with $f_{\rm{HI}}b~=~0.012~\pm~0.003$ as measured by \citet{chang_et_al_2010}.  To obtain individual constraints on these parameters, one will need to measure
redshift-space distortions themselves.

\subsection{Sensitivity of an Array to the 21cm Signal}
\label{sec:sensitivity}

There are three independent sources of statistical uncertainty in a 21cm power spectrum measurement: thermal
noise in the interferometric visibilities, sample variance, and shot noise, of which the last is in some
sense ``signal," but still inhibits measurements of cosmological parameters.
For the first-generation of 21cm experiments, thermal noise will be the dominant source of
uncertainty in measurements of the power spectrum.  We therefore calculate the effects
of thermal noise independently in \S\ref{sec:thermal_noise}.  We add in the effects of sample
variance in \S\ref{sec:sample_variance} and argue in \S\ref{sec:shot_noise} that shot noise
can be neglected for these experiments.

\subsubsection{Thermal Noise}
\label{sec:thermal_noise}
Thermal noise, in addition to being the dominant source of uncertainty in first generation
21cm BAO experiments, is also likely to be
least familiar to those used to optical redshift surveys.
Given the limited collecting area of early experiments, reducing thermal noise
contributions is of paramount importance, even at the expense of the number of
Fourier modes measured.
Much of the work in this section explicitly follows the derivation of an interferometric array's
power spectrum sensitivity presented in P12a (only the prefactors have changed to account for
different fiducial values at $z=1$).  The approach of P12a is to treat each baseline
as an independent probe of the 21cm power spectrum at one $k_{\perp}$ set by the baseline
length, and many $k_{\parallel}$-values along the line-of-sight.  We therefore first derive
the sensitivity of a single-baseline; the sensitivity of the array is the aggregate sum of
all these independent measurements.

We begin with a version of 
equation 16 from P12a, which gives
the power spectrum of the thermal noise obtained from one integration of a
dual-polarization baseline:
\begin{equation}
\label{eq:single_baseline_sensitivity}
\Delta^2_{\rm N}(k) \approx X^2Y\frac{k^3}{2\pi^2}\frac{\Omega}{2t}T_{\rm sys}^2,
\end{equation}
where $X^2Y$ is a scalar translating observed units to cosmological distances in $h^{-1}$Mpc
($X$ converts from angles on the sky to transverse distances, $Y$ from bandwidth to
line-of-sight distance), $\Omega$ is the solid angle of the primary beam of one element
in steradians, $T_{\rm sys}$ is the system temperature, and $t$ is the amount of
integration time of the sample.  It is again worth emphasizing
that this equation is not valid for all $k$-modes (which would imply a white noise power spectrum
throughout Fourier-space), but rather only those modes sampled by the one baseline in question.

We also note that this relationship is very similar to equation 31 in \citet{ansari_et_al_2012b}.
Quantitatively, however, they differ by a factor of 4.  There are two separate effects contributing
to this difference.  Firstly, our equation is explicitly for a dual-polarization receiver, giving
us twice as many independent samples of the same $k$-modes and therefore half the noise.  Secondly,
for the receiver design of the BAOBAB system, the RMS fluctuations in a measurement are given by
$T_{\rm rms} = \frac{T_{\rm sys}}{\sqrt{Bt}}$ \citep{kraus_1966,thompson_et_al_2001}.  
\citet{ansari_et_al_2012b} include an additional factor of 2 in their equation (21),
perhaps due to the design of their system including phase switching or some other effect.

The science goals of a BAO experiment are to actually measure $X$ and $Y$; that is, since
the exact values of $X$ and $Y$ depend on the underlying cosmology, we can combine the
known physical scale of BAO with the angular and frequency scales in the observed signal to extract
the detailed expansion history of the universe.  For the purpose of a sensitivity derivation,
however, the behavior of $X$ and $Y$ can be considered well enough known to compute fiducial
values for equation \ref{eq:single_baseline_sensitivity}.

$X$ is related to the angular size distance, $D_{\rm A}$, as
\begin{equation}
\label{eq:X}
    X=D_{\rm A}(1+z)\equiv\int_0^z{\frac{c\ dz}{H(z)}},
\end{equation}
with $H(z)$ in the matter/dark-energy dominated epoch being approximately given by 
$H(z)=H_0\sqrt{\Omega_{\rm M}(1+z)^3+\Omega_\Lambda}$.  Numerical integration for a flat universe
with $\Omega_{\rm M}=0.27$, $\Omega_\Lambda=0.73$, and $H_0=70$ yields $D_{\rm A}(z=1)\approx1680$ 
proper $\rm{Mpc}$ \citep{wright2006}.  Ignoring the evolution in the 
angular diameter distance around $z\sim1$, we can write:
\begin{equation}
\label{eq:X_approx}
    X\approx 1700 (1+z) \frac{\rm Mpc}{\rm radian}.
\end{equation}
Note that although we use this (admittedly somewhat poor) approximation 
for simplicity
in deriving the relations for thermal noise
power spectra in this subsection, all the subsequent results include the full $z$-dependence of the
angular diameter distance.

A few more words are warranted concerning the magnitude of $X$ at $z=1$.  Given the scaling of equation \ref{eq:X},
a $16\lambda$-baseline (6.8m a $z=1$) corresponds to a transverse wavemode of 
$k_\perp = 0.042~h\rm{Mpc}^{-1}$, a non-negligible value compared to the first BAO peak at 
$\sim 0.08~h\rm{Mpc}^{-1}$.  Therefore, baselines longer than $\sim 32\lambda$ will 
lose access to the first peak and be less effective
probes of cosmology, regardless of foreground effects to be discussed later.  This scaling
motivates the extremely compact configurations proposed for BAOBAB in \S\ref{sec:baobab}, despite
the possible systematics associated with such short 
baselines.\footnote{Epoch of Reionization experiments at $z=9$, however, do not find themselves 
limited by the transverse modes probed. $X(z=9) \approx 9360 \frac{\rm Mpc}{\rm radian}$ (P12a), 
so that a $16\lambda$-baseline corresponds to $k_{\perp} = 0.015~h\rm{Mpc}^{-1}$.  
With this scaling, the effect of the $k_{\perp}$ component on the measured power spectrum will 
always be sub-dominant to the foreground effects described in P12b.}

To compute the scaling between frequency, $\nu$, and comoving line-of-sight distance, $r_{\rm los}$, we use
\begin{equation}
    dr_{\rm los} = \frac{c dz}{H(z)}.
\end{equation}
Since $\nu(1+z)=\nu_{21}$, we have that $dz/(1+z)=-d\nu/\nu$, so
\begin{equation}
\label{eq:Y}
    Y\equiv\frac{dr_{\rm los}}{d\nu}=\frac{c(1+z)^2}{\nu_{21} H(z)}.
\end{equation}
Evaluating the above numerically, we get
\begin{equation}
    Y = 3.0\frac{(1+z)^2}{\sqrt{\Omega_{\rm M}(1+z)^3+\Omega_\Lambda}}\ \frac{\rm Mpc}{\rm MHz}.
\end{equation}
Finally, we compute the product $X^2Y$ used in equation \ref{eq:single_baseline_sensitivity}:
\begin{equation}
    X^2Y\approx2.93\frac{(1+z)^4}{\sqrt{\Omega_{\rm M}(1+z)^3+\Omega_\Lambda}}\ \frac{{(h^{-1}\rm Mpc})^3}{{\rm sr}\cdot{\rm Hz}}.
\label{eq:cosmo_scaling}
\end{equation}
Nominally, $X^2Y=28 (h^{-1}\rm{Mpc)^3~sr^{-1}~Hz^{-1}}$
 at $z=1$.\footnote{The magnitude of $X^2Y$ at $z=1$ represents 
an often under-appreciated gain in sensitivity between a 21cm BAO experiment and a similar
reionization experiment.  At $z=9$, $X^2Y \approx 540(h^{-1}\rm{Mpc)^3~sr^{-1}~Hz^{-1}}$, 
meaning that the quoted EoR signal strength of
$\sim10\rm{mK}^2$ is normalized over a much larger volume.  The smaller volume scalar at $z=1$
means that over an order of magnitude less thermal noise is picked up per equivalent integration.}

The other values in equation \ref{eq:single_baseline_sensitivity} are system-dependent parameters.
The BAOBAB tiles have a considerably sized primary beam on the sky, $\Omega \approx 150 \rm{sq. deg}\approx 0.045$sr.\footnote{This value for primary beam was estimated based on models of the PAPER
single dipole primary beam and a simple array-factor to estimate a tile beam.  The actual value
will depend on the illumination pattern of the ground screen flaps by the dipole.  If prototype
systems substantially under-perform in this regard, a second-stage of element design may be necessary
to reduce the size of the primary beam in order to achieve the sensitivities presented here.}
However, this beam is significantly narrower than that of a single dipole,
so that the use of beamforming results in a considerably lower noise level,
since our noise power spectrum scales as $\Omega$ into equation \ref{eq:single_baseline_sensitivity}.

The issue of system temperature is another instance where a BAO experiment at $z=1$ is
fundamentally different from an EoR experiment at $z=9$.  In the EoR case, Galactic synchrotron emission is the primary source of noise at 150 MHz,
with a value of $\sim300$K toward the galactic poles.  However, synchrotron emission scales approximately as
\begin{equation}
    T_{\rm sync}\approx 300 {\rm K} \left(\frac{\nu}{150 {\rm MHz}}\right)^{-2.5}.
\end{equation}
At frequencies around $750$ MHz, $T_{\rm sync}\approx 5{\rm K}$.  This value is substantially below
typical receiver temperatures of $50{\rm K}$, so that receiver temperatures will dominate $T_{\rm sys}$.  As described
in \S\ref{sec:baobab}, BAOBAB will have a system temperature of approximately 50K across its
entire band.

We can express the noise power spectrum of one dual-polarization 
baseline integrating on one Fourier mode 
by substituting these fiducial values in equation \ref{eq:single_baseline_sensitivity}.
However, before doing so, it is worthwhile to look ahead and emphasize that the remaining
results in this section are intended to give both a quantitative sense of the level of
thermal noise in $z\sim1$ 21cm observations and scaling relations for the effect of various
instrumental and observational parameters.  In the sensitivity calculations of \S\ref{sec:baobab32} and \S\ref{sec:baobab128}, however, we fully simulate the $uv$-coverage of our arrays, including
the effects of earth-rotation synthesis.  We then use equation \ref{eq:single_baseline_sensitivity} to
evaluate the thermal noise level in each $uv$-pixel given an effective integration time in that pixel.
 
Substituting our fiducial values into equation \ref{eq:single_baseline_sensitivity} yields the following result for the sensitivity of a single baseline to the 21cm power spectrum:
\begin{align}
\label{eq:quant_single_baseline_sensitivity}
    \Delta^2_{\rm N}(k) \approx &~8\times10^{-3} \frac{(1+z)^4}{\sqrt{\Omega_{\rm M}(1+z)^3+\Omega_\Lambda}}
    \nonumber\\&\times
    \left[\frac{k}{0.1\ h\rm{Mpc}^{-1}}\right]^3
    \left[\frac{\Omega}{0.045~\rm{sr}}\right]^\frac32
    \nonumber\\&\times
    \left[\frac{180~\rm{days}}{t_{\rm days}}\right]
    \left[\frac{|\vec{u}|}{20}\right]
    \left[\frac{T_{\rm sys}}{50~\rm{K}}\right]^2 \rm{mK}^2,
\end{align}
where $|\vec{u}|$ is the length of the baseline in wavelengths.
We have also replaced $t$, the integration time in a sample, with a combination of factors of 
$|\vec{u}|$ and $\Omega$.  These quantities are related to the amount of time a single-baseline
will sample a $uv$-pixel:
$\Omega^\frac12$ sets the diameter of the pixel
and $|\vec{u}|$ sets the length of time the baseline samples that same pixel
before earth rotation causes it to move into an independent pixel (longer
baselines drift through the $uv$-plane faster).  
The full derivation of this relation is given in \S2.2 of P12a.
The choice of a 20-wavelength baseline
is arbitrary.  
We have also added a $t_{\rm days}$ factor; each day of an observation provides an identical measurement to the
previous day, resulting in a linear increase in the power spectrum sensitivity.
Our fiducial
``long observation" is 180 days; we set this value as a hard maximum for the number
of days BAOBAB can observe in one calendar year.  This choice comes from the fact that observations
will be compromised by foreground emission when either the Galactic plane or the sun is in view.

In order to calculate the sensitivity of an entire interferometric array, we must sum
the contributions of all the Fourier modes sampled by every baseline, 
paying careful attention to the number of times a unique Fourier mode is sampled by distinct
redundant baselines.
If each baseline measured an independent Fourier mode, the overall 
power spectrum sensitivity of an array would grow 
proportionally to the square root of the number of baselines (or, alternatively, linearly
with number of antennas).
To first approximation, every baseline can contribute an independent measurement of each $|k|$-mode
in the power spectrum, since the frequency axis covers a very broad range in $k$.
\footnote{
The situation is more complicated than this for two reasons.  
First, each baseline has a minimum $|k|$-mode it can measure, corresponding
to its length in $k_{\perp}$.  The analytic formulae below ignore this effect,
which becomes important for small values of $k$.  However, we emphasize that the cosmological
sensitivity results in the subsequent sections do take this effect into account.
Second, redshift space distortions break the isotropy between $k_{\parallel}$ and 
$k_{\perp}$ so that modes of equal $|k|$ cannot na\"{i}vely be combined.
As stated previously,
we maintain 2D information in our full analysis and only combine modes in annuli of equal 
$k_{\perp}$.  Only for the purposes of making 1D power spectrum plots, do we combine
all modes of equal $|k|$ by reducing the noise in each mode by $(1 + \beta\mu^2)^2$.
}
However, coherent integration on a particular Fourier mode beats down thermal noise as the 
square-root of the number of samples in \emph{temperature}, and hence linearly in the temperature-squared
units of a power spectrum.  Therefore, redundant measurements can improve the power
spectrum sensitivity of an array to select Fourier modes faster than two non-redundant baselines 
measuring independent modes of the same magnitude.
In our formalism, this additional sensitivity boost from redundant sampling 
enters through the $f/f_0$ metric for the amount of redundancy in an array configuration, 
defined in P12a as:
\begin{equation}
\frac{f}{f_0} \equiv \frac{\sum\limits_i n_i^2}{\sum\limits_i n_i},
\end{equation}
where $i$ labels individual $uv$ pixels, and $n_i$ the number of one-second integration samples
falling within pixel $i$.
The ratio $f/f_0$ measures the increase in sensitivity 
for a redundant array over one in which there is no sampling
redundancy from either redundant baselines or redundant time samples.
(This hypothetical ``reference" array is obviously non-physical, as it assumes each independent
integration in time will measure a statistically independent sky; in practice, this means that the
$f/f_0$ metric incorporates the length of time an array can observe the same patch of sky.)   
An $f/f_0$ factor of $10^4$ is representative of the 32-element 
drift-scanning maximum-redundancy arrays described in P12a.

Using this metric, we can express the resultant sensitivity of an arbitrary array as:
\begin{align}
\label{eq:noise_pk}
    \Delta^2_{\rm N}(k)\approx &~2 \times 10^{-4}
    \frac{(1+z)^4}{\sqrt{\Omega_{\rm M}(1+z)^3+\Omega_\Lambda}}
    \nonumber\\&\times    
    \left[\frac{k}{0.1\ h\rm{Mpc}^{-1}}\right]^3
    \left[\frac{\Omega}{0.045\ {\rm sr}}\right]
    \left[\frac{T_{\rm sys}}{50\ {\rm K}}\right]^2
    \nonumber\\&\times    
    \left[\frac{8\ {\rm hrs}}{t_{\rm per\_day}}\right]^\frac12
    \left[\frac{180\ {\rm days}}{t_{\rm days}}\right]
    \left[\frac{32}{N}\right]
    \left[\frac{10^4 f_0}{f}\right]^\frac12
    {\rm mK}^2.
\end{align}
This equation is derived in Appendix B.2 of P12a.
The $f/f_0$ term is computed from $uv$ sampling patterns including earth
rotation aperture synthesis effects over a period of one hour phased to a single pointing center.
One hour roughly corresponds to the time it takes a point on the sky to drift through the
width of the BAOBAB primary beam, after which 
a statistically independent patch of sky comes to dominate the data,
so that additional integration time only grows the sensitivity as $(t_{\rm per\_day})^\frac12$.
In general, $f/f_0$ accounts for most
all effects regarding array configuration, so that the additional factors of $\Omega^\frac12$
and $|\vec{u}|$ that appeared in equation \ref{eq:quant_single_baseline_sensitivity} do not
appear in \ref{eq:noise_pk}.  
The factor of $t_{\rm per\_day}$ sets the total integration time per day, which will likely
be limited by the size of a low foreground emission region (i.e., a ``cold patch").  We choose
8 hours as the maximum time that can spent observing cold patches per day, a value influenced
by existing all-sky maps and observations with PAPER.

\subsubsection{Sample Variance}
\label{sec:sample_variance}

In galaxy redshift surveys, the sample variance can be calculated relatively simply by 
counting Fourier modes over an effective survey volume.  However, in the case of BAOBAB,
not all modes are equal, since we have used redundant samples of certain modes to beat down thermal
noise.  In creating a 1- or 2-D power spectrum out of the full 3-D Fourier space,
we must take a weighted combination of these modes, since the ratio of 
thermal noise to sample variance can vary between every $k$-mode measured.

Using inverse-variance weighting to combine each measurement at a particular $k$-mode, 
one finds that the optimal estimator of the power spectrum results in an error
that can be calculated by combining the errors on each measured mode in inverse-quadrature:
\begin{equation}
\label{eq:mode_sum}
\delta \Delta^2(k) = \left(\sum\limits_i\frac{1}{\left(\Delta^2_{\rm{N},i}(k)+\Delta^2_{21}(k)\right)^2}\right)^{-\frac12},
\end{equation}
where $\delta\Delta^2(k)$ is the resultant uncertainty on a given $k$-mode, $\Delta^2_{{\rm N},i}$ 
is the per-mode thermal noise calculated with equation \ref{eq:single_baseline_sensitivity}
taking the full $uv$-coverage and earth-rotation synthesis into account, $\Delta^2_{21}(k)$
is the cosmological 21cm power spectrum (which is also the sample variance error), and $i$
is an index labeling the independent $k$-modes measured by the array over which we are summing
(note that we never combine modes into bands with significantly varying 
$|\vec{k}|$, which is why we can sum inverse absolute variance instead of 
inverse fractional variance).

\begin{figure}\centering
\includegraphics[width=3.5in]{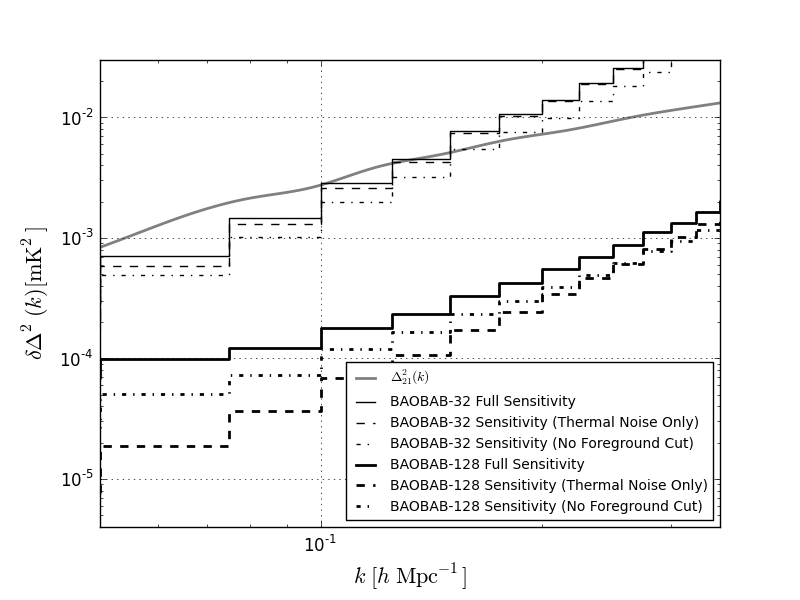}
\caption{Noise levels for two fiducial observations of one declination range 
with BAOBAB at $z = 0.89$.  
The dashed thin (thick) line shows
the sensitivity given thermal noise only for a 30 days observation with a 
35-element system
(180 days with a 132-element system); 
solid lines show the effect of including sample variance.  
We also plot the effect of our foreground model of \S\ref{sec:dspec}; 
the achievable sensitivity if no modes
are excluded by foregrounds is shown by the dot-dashed lines.
The other two curves do include the effects of the foreground model.
For comparison, we also plot the dimensionless 21cm power spectrum as the thick, gray line.
At the larger scales, the 180 day observation with 132-tiles is completely 
dominated by sample
variance.
At the smaller scales, the analytic expression for thermal noise
given in equation \ref{eq:noise_pk} accurately reproduces the sensitivity.  
The plotted thermal noise
only curve is not a perfect power-law because longer baselines cannot probe the largest scale
$k$-modes. 
} \label{fig:sense_wwo_samp_var}
\end{figure}

Since the sample variance is completely a function of the $uv$-coverage of an array,
it is best calculated
numerically, as described in the preceding section.  In Figure \ref{fig:sense_wwo_samp_var} we plot the 1D thermal noise and 
sample variance uncertainties for
two maximum-redundancy configurations of BAOBAB --- 35- and 132-tiles ---
shown in Figures \ref{fig:configuration} and \ref{fig:antpos128}.
To calculate the sample variance,
we use the 21cm brightness power spectrum from CAMB, also plotted for comparison.
At the scale of the first acoustic peak, sample variance has clearly become the dominant source 
of error for a long integration with 132-elements; we discuss possible methods for improving this
situation in \S\ref{sec:baobab128} and \S\ref{sec:improving}.  At the smaller scales, however, the analytic functions given in
the previous section accurately reproduce the array sensitivity.  Note that the thermal noise
only curve in Figure \ref{fig:sense_wwo_samp_var} is not a perfect power law because not all 
baselines can probe the largest scales. 
We also plot the effect of our foreground model of \S\ref{sec:dspec}; 
the achievable sensitivity if no modes
are excluded by foregrounds is shown by the dot-dashed lines.
The other two curves do include the effects of the foreground model.

\subsubsection{Shot Noise}
\label{sec:shot_noise}

Measurements of the 21cm power spectrum will also be affected by the discrete nature of
the neutral hydrogen distribution at low redshift.  Only overdensities self-shielded to the $\sim1$ Ry ionizing background
contain neutral hydrogen, so we will be subject to the same galactic shot noise as
optical redshift surveys.  Following \citet{seo_et_al_2010}, we assume that galaxy
positions and luminosities are distributed with probability proportional to
$1+b~\delta_m$, where $b$ is the bias and $\delta_m$ is the mass overdensity.  This allows
us to treat shot noise as a scale-independent contribution to the power spectrum
with $P(k) = 1/\bar{n}$.  In terms of 21cm brightness:
\begin{align}
\Delta^2_{\rm shot}(k) \approx \left[\tilde{T}_{21}(z)\right]^2 \frac{1}{\bar{n}} \frac{k^3}{2\pi^2}
\end{align}
Using the result of \citet{seo_et_al_2010} (who use a fit to the neutral hydrogen mass
function of \citealt{zwaan_et_al_2005}), $\bar{n} = 0.01~h^{3}\rm{Mpc}^{-3}$ and 
$\Delta^2_{\rm shot} \approx 6.5 \times 10^{-5} \rm{mK}^2$ at $k = 0.1~h\rm{Mpc}^{-1}$ and 
$z = 1$.  The number density of hydrogen-containing halos is substantially higher than for
the bright galaxies used in optical/NIR surveys, making shot noise a substantially smaller 
contaminant; it will only begin to dominate the signal
at $k > 2~h\rm{Mpc}^{-1}$.
Regardless of uncertainties in this calculation, shot noise is clearly
a subdominant effect, and we neglect it for the remainder of this work.

\subsection{The Delay Spectrum Technique at $z\sim 1$}
\label{sec:dspec}

Before combining the results of the last two sections, we must discuss the effect of
foregrounds on power spectrum measurements.  The presence of foreground emission orders
of magnitude brighter than the cosmological 21cm signal has been one of the major impediments
for high-redshift 21cm tomography.  P12b presented a per-baseline delay-spectrum
technique for isolating foreground emission solely on the basis of its spectral smoothness.
In this section we briefly recapitulate the principles of the delay-spectrum technique, and present
a simple approximation for the behavior of foregrounds in the 600-900 MHz band.

The delay spectrum technique is a methodology for using each baseline of
an interferometer as a independent probe of the 21cm power spectrum.  The most powerful
aspect of this approach is that the frequency dependence of a baseline's Fourier
sampling pattern, typically regarded as a major complication for 21cm experiments,
naturally gives rise to an isolation of foreground emission in Fourier space.  The ability
to remove foregrounds on a per-baseline basis allows multiple baselines to be tuned to
target the same Fourier mode for greater sensitivity, as opposed to more traditional
techniques that use overlapping $uv$-coverage at multiple frequencies to avoid the issue
of frequency-dependent sampling.

At the heart of the delay transform is a dual interpretation of the Fourier transform
of interferometric visibilities along the frequency axis.  On the one hand, for 21cm
experiments, frequency maps directly into redshift since the observed signal is a spectral
line.  Therefore, the Fourier transform along the frequency axis gives $k_{\parallel}$, the
Fourier wavemode along the line of sight.  However, the frequency dependence of a baseline's
length (as measured in wavelengths), gives rise to the delay transform interpretation of the
frequency Fourier transform presented in \citet{parsons_and_backer_2009}.  
If performed over a wide enough bandwidth, this transform
maps sources to Dirac delta functions in ``delay space," corresponding to the geometric
delay of signal arrival time between the two elements of the baseline.  There is thus
a maximum delay at which any signal coming from the sky can appear, set by the physical
length of the baseline.  Furthermore, each source delta-function will be convolved
by a kernel representing the Fourier transform of that source's intrinsic spectrum (as
well as any spectral features introduced by the instrument).  Therefore, as long as the
instrumental frequency structure is kept to a minimum, sources with smooth intrinsic
spectra (such as foreground emission) will have their emission confined within the
region of delay space set by the maximum delays (the so-called ``horizon-limit").  
Sources with unsmooth emission,
like the 21cm signal, will be convolved by a broad kernel, scattering ``sidelobes"
well beyond the horizon limit, and creating a window for detecting 21cm emission free of
smooth-spectrum foregrounds. 

A major component of P12b was to calculate the mapping between cosmological $k$-space and
delay space.  To phrase the same question in other terms, we explicitly calculated the effect
of ``mode-mixing" due to the frequency dependence of a baseline's $k_{\perp}$ sampling
on the recovery of the 21cm power spectrum.  For short baselines like those used in BAOBAB,
delay-modes proved an effective probe of the 21cm power spectrum, recovering the signal
without corruption due to mode-mixing. 

A full simulation quantifying the effects of the delay transform on foregrounds
is beyond the scope of this work.  Rather, we assign a minimum 
$k_{\parallel}$ which depends on baseline length, 
below which we consider modes as being wholly contaminated by foregrounds.
These contaminated modes are treated as not ``measured" by the array, that is, they are
excluded from the sum in equation \ref{eq:mode_sum}.
Since the horizon limit described above is a linear function of baseline length, we 
use $k_{{\parallel},{\rm min}}$ which linearly increases on longer baselines.
We model our choice for the exact value of $k_{{\parallel},{\rm min}}$ on the simulations presented for PAPER
in P12b, which finds for $16\lambda$-baselines at 150~MHz
foregrounds contaminate modes with $k_{\parallel} \lesssim 0.2~h\rm{Mpc}^{-1}$.  
At EoR frequencies of 150 MHz, this cutoff maps to delay-modes of 400 ns.  Since BAOBAB
baselines are physically shorter by a factor of 5, this reduces the maximum
delay-space contamination to 80 ns, which in turn maps back to $k = 0.1~h\rm{Mpc}^{-1}$
at 750 MHz, using the $Y$ parameter from equation \ref{eq:Y}.
There are two important factors which will further serve to reduce this number
for BAOBAB.  First, celestial foregrounds should have power law spectra with steeply decrease
in intensity versus frequency, and so will be fainter than at EoR frequencies. 
Although the signal has also fallen a similar amount, this reduced foreground structure will still make the delay transform even more
effective at isolating foreground emission.  Secondly, the narrower primary beam of BAOBAB
will limit the delay modes from which there can be appreciable celestial emission,
as sources near the horizon will be significantly attenuated.  

To better determine the scale of $k_{{\parallel},{\rm min}}$ in our foreground model,
we perform a cursory calculation in which the delay-transform is applied to a simulated sky model.  
In these simulations, we assume the sky is entirely composed of point sources, where the source
strength distribution follows a power-law with a slope of -2.0, normalized
to a 2 Jy source per 10 steradians, a distribution derived empirically
from PAPER data with extrapolation to the BAOBAB band.  
We also model the frequency spectrum of each
source as a power-law with a normal distribution of spectral indices centered on -1.0
and a standard deviation of 0.25.  We refer to these
simulations as ``cursory," since they exclude instrumental effects such as RFI flagging
and frequency-dependent beam structure.  
Instead, we use a single, frequency-independent Gaussian to model the primary beam of the
BAOBAB tile; the potential effects of a more realistic beam model are discussed in \S\ref{sec:discussion}.
We find the delay transform confines foregrounds to $k$-modes below a value of 
$k_{\parallel} = 0.045~h{\rm Mpc}^{-1}$ for baselines of 16 wavelengths.
The $k_{\parallel}$ value for the maximum delay of a 16$\lambda$ baseline at 
750 MHz (i.e., the 
horizon limit) is $0.028~h{\rm Mpc}^{-1}$, which implies that the intrinsic spectral behavior of 
foreground emission
corresponds to a kernel of width $\sim 0.02~h{\rm Mpc}^{-1}$.  
In this work, our foreground model
is to exclude $k$-modes smaller than the sum of the maximum realizable delay on a baseline
(converted from seconds of light-travel time to $h{\rm Mpc}^{-1}$ using equation \ref{eq:Y}) 
and this kernel.  
The maximum realizable delay scales linearly with baseline length, while the additive kernel 
remains constant.
In effect, this model states that intrinsic spectral structure in foregrounds 
corrupt $0.02~h{\rm Mpc}^{-1}$ beyond a na\"{i}ve prediction based only on the physical 
length of the baseline.  We explore the effects of modifying this model in \S\ref{sec:shortcomings}.

\subsection{Detecting the HI Power Spectrum}
\label{sec:baobab32}

The first major science result from BAOBAB will be the detection of the 21cm power spectrum near 
$z\sim1$.  
We present predictions for the power spectrum error bars using the formalism outlined above:
we fully simulate the $uv$-coverage of our arrays, including
the effects of earth-rotation synthesis, over a one-hour period.  
We then use equation \ref{eq:single_baseline_sensitivity} to evaluate the thermal noise level in 
each $uv$-pixel given an effective integration time in that pixel.  
We incorporate redshift-space distortion effects by reducing the magnitude of these thermal
noise errors according to equation \ref{eq:rsd}.
We include the effects of sample variance in our measurements using equation \ref{eq:mode_sum}, 
combining measurements with the same $|k_{\perp}|$, but maintaining 2D information in the
($k_{\perp},k_{\parallel}$)-plane.  
These error bars are further reduced by both the square root of the number of 
independent 1-hour pointings available per-day (8 in our fiducial calculation)
and by the number of days observed.
For the plots below, we further compress these errors into
1D, but use the full 2D information for our calculations of detection significance and 
cosmological parameter extraction in \S\ref{sec:baobab128}.

\begin{figure}\centering
\includegraphics[width=3.5in]{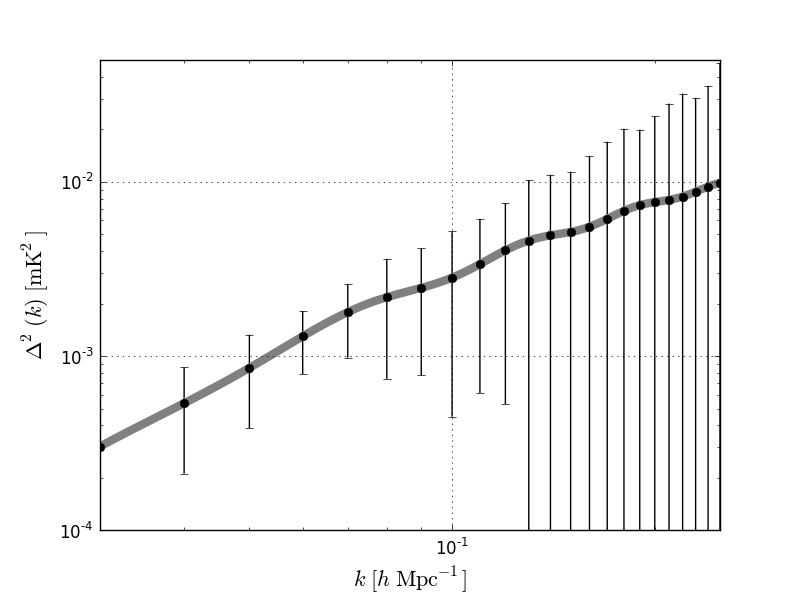}
\caption{Predicted constraints on $z = 0.89$ 21cm power spectrum from a 30 day observation with
a 35-tile BAOBAB system.  The net result is a $\sim5\sigma$ detection of the
power spectrum.  Results are comparable in the other 2 redshift bins. 
} \label{fig:baobab32_sense}
\end{figure}

A high significance detection will be achievable with a short $\sim1$-month observation
with a 35-tile system operating in the 
maximum redundancy shown in Figure \ref{fig:configuration}.\footnote{Correlator inputs have 
traditionally been in powers of 2; hence this array is our $\sim 32$-tile configuration. The sensitivity calculations do assume that all 35 elements are correlated.}  
The predicted measurement for a 30-day observation 
(240 hours) is shown in Figure \ref{fig:baobab32_sense}.  
These observations assume a 100-MHz
bandwidth centered on 750 MHz ($z = 0.89$).  The net result is an $5.6\sigma$ detection of
the 21cm power spectrum when our model for foreground emission from \S\ref{sec:dspec}
is used to exclude contaminated modes.  
Results for bands centered on $650~(z = 1.18)$ and 850 MHz $(z = 0.67)$
are similar, yielding $5.8\sigma$ and $5.0\sigma$ detections, respectively.
Although a small effect, we do modify the system temperature in each band to represent
the change in sky temperature; a spectrally-flat $T_{\rm sys}$ is therefore only assumed 
on 100~MHz scales.
Rather, the lower
significance detection at the lowest redshifts results primarily from the scaling of the angular diameter
distance; at redshift $z = 0.67$, a $16\lambda$ baseline corresponds to a $k_{\perp}$ wavemode
of $\sim0.06~h{\rm Mpc}^{-1}$, limiting the number of baselines that can probe
the largest-scale $k$-modes where thermal noise is lowest.  
Over the $z = 0.5-1.5$ range, both the 21cm signal and,
the noise remain roughly constant in magnitude, the latter because it is dominated 
by a frequency-independent
front-end amplifier noise temperature.  This trend does not continue indefinitely, however,
as sky noise increases with increasing redshift, eventually dominating the system temperature.
  
Measurements of this significance will allow for an accurate determination of $f_{\rm HI}~b$, 
the combination of the cosmic neutral
hydrogen fraction and the bias of neutral hydrogen containing regions, as a function of redshift.  Breaking the degeneracy between these parameters will require
additional information.  Measuring redshift-space distortions can, in principle,
separate the effects of the two terms.
Constraints from a longer integration or a system with $\sim 64$ elements
will further improve constraints on the neutral hydrogen power spectrum, and it will be possible
for these systems to measure redshift-space distortion effects.  
Measuring 
these effects accurately requires more careful systematic control, which may
warrant different configurations and observing strategies, so we postpone an 
exploration of this science to a future work (see, e.g., \citealt{masui_et_al_2010}).

\subsection{Detecting Baryon Acoustic Oscillations}
\label{sec:baobab128}

\begin{figure}\centering
\includegraphics[width=3.5in]{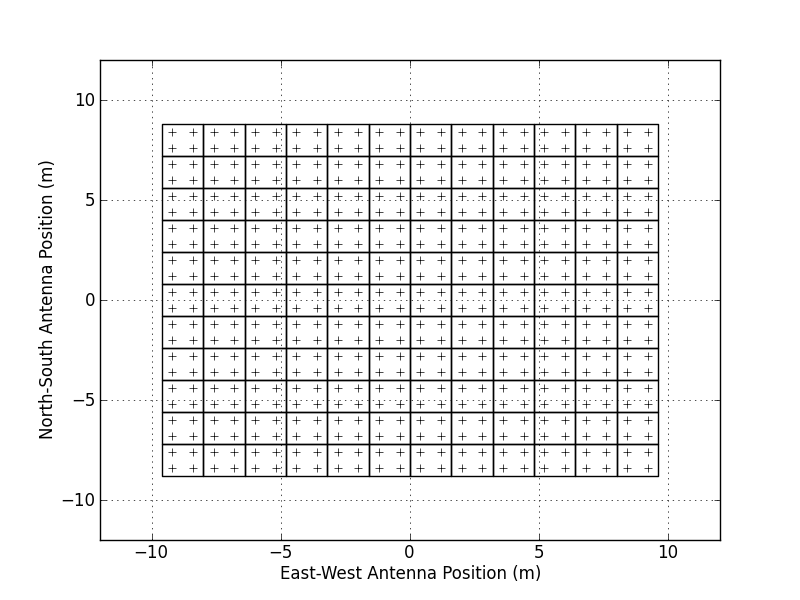}
\caption{The maximum redundancy configuration of a 132-element BAOBAB system.  The close-packed
tiles are chosen to produce the shortest possible baselines.
} \label{fig:antpos128}
\end{figure}

As shown in in Figure \ref{fig:sense_wwo_samp_var}, a 132-tile BAOBAB array with the configuration
shown in Figure \ref{fig:antpos128} has effectively reached
the sample variance limit in 180 days (1440 hours) of observing time.  
Using the methodology described at the beginning of \S\ref{sec:baobab32},
we calculate that this observation yields a $3.3\sigma$ detection of the BAO features at $z = 1.18$, 
with effective $2.1-$ and $2.7\sigma$ non-detections at the $z = 0.67$ and $z = 0.89$ bands,
respectively, where we have isolated the BAO features from the broad-band shape of the
power spectrum by removing a model fit using the transfer function from
\citet{eisenstein_and_hu_1998}.
The effect of sample variance is most dominant at the lower redshifts, 
because the angular diameter distance scaling means that fewer samples of the BAO scale can be
found in the same area of sky (this observation of 8-independent fields with a
0.045 steradian primary beam corresponds to an effective survey area of $\sim 1200$ square degrees).
While longer observations with the same array configuration can improve these constraints
by reducing thermal noise on the smaller scale modes, a
better approach will be to observe additional independent fields.

For our fiducial BAOBAB observation, we use an array which observes 24 independent fields
(i.e., three independent declination observations for 8 hours per day), yielding an
effective survey area of $\sim 3600$ square degrees.  
We discuss the motivation for this particular approach to increasing survey volume in
\S\ref{sec:improving}.
There are several equivalent ways an 
experiment can probe this additional area.  One approach would be to conduct three 1-year surveys,
with the dipoles pointed towards a different declination; this could be achieved by physically
placing the array on an platform inclined by $\sim 15^{\circ}$, or potentially by adding a steering
component to the tile-beamformers.  If the beamformers are designed to allow multiple beams, one
could in principle achieve similar sensitivities with only one year of observing, although
at the expense of additional degrees of complexity in the system.  If funding permits, the 
simplest approach might be to build three 132-tile BAOBAB arrays, each tilted towards a different
declination; this would also yield the subsequently predicted sensitivities in only one season of
observing.  Note that for BAO science these independent configurations are
potentially more desirable than
an array with a larger number of fully correlated tiles.  Since we are using a close-packed
configuration, the addition of more tiles can only yield new modes at corresponding larger $k_{\perp}$ 
where the amount of BAO information is significantly diminished.  It may be possible
that more information could be recovered from these larger $k$ modes using a reconstruction
method (e.g. \citealt{eisenstein_et_al_2007}, \citealt{padmanabhan_et_al_2009},
\citealt{noh_et_al_2009}), but we do not explore this option in this work.  
If there is significant BAO information beyond $k \sim 0.2$, then a $\sim 256$-tile array could
possibly yield tighter constraints than two 132-tile arrays observing for the same
amount of time.
In \S\ref{sec:improving},
we discuss other ways to increase the survey footprint and further reduce the effect of sample
variance.

\begin{figure}\centering
\includegraphics[width=3.5in]{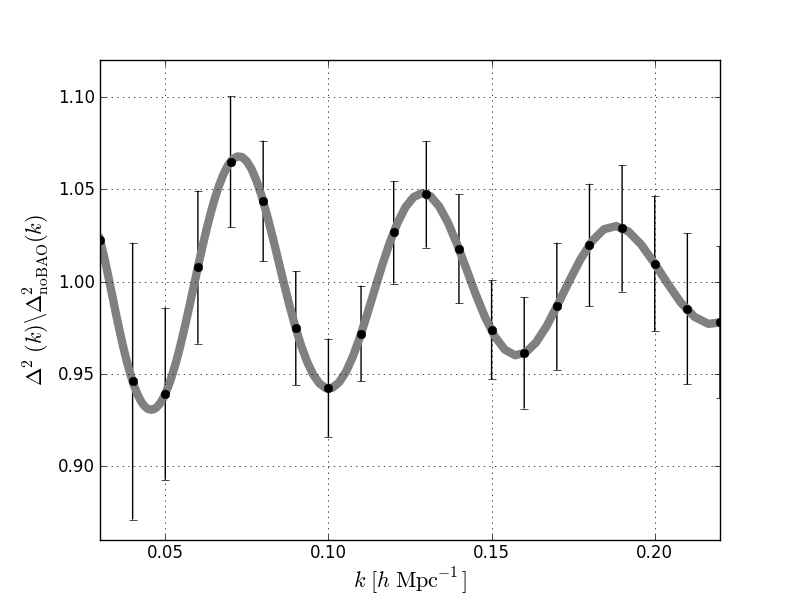}
\caption{Predicted constraints on the $z = 0.89$ BAO features from a 180 day observation
of three-declination fields with
a 132-element BAOBAB system.  The net result is a $4.7\sigma$ detection of the BAO features.  The sensitivity is comparable in the other 2 redshift bins.
} \label{fig:baobab128_sense}
\end{figure}

We plot the expected constraints for the three-declination range fiducial 
BAOBAB observations on
the $z = 0.89$ BAO features in Figure \ref{fig:baobab128_sense}, where the 
broad band
shape of the power spectrum has been removed using the transfer function fit
of \citet{eisenstein_and_hu_1998}, which neglects BAO.  
The measurements from a 180-day integration at each declination range with this array 
amount to an 4.7$\sigma$ detection of these features.  Results are similar
for the other redshift bins, with expected 3.6$\sigma$ and 5.7$\sigma$ detections at redshifts of 
0.67 and 1.18, respectively.  While it is clear from Figure \ref{fig:sense_wwo_samp_var} that
sample variance dominates the errors on the largest scale modes after 180 days of observing
one declination range, we find the additional sensitivity towards the higher BAO peaks with this
observing duration yields better constraints on the signal than e.g., observing twice as many
declination ranges for 90 days.  

With a significant BAO detection, BAOBAB can also begin to place 
constraints on cosmological parameters.  To quantify the effect of such measurements,
we use the Fisher matrix formalism of the Joint Dark Energy Mission (JDEM) Figure of Merit
Science Working Group (FoMSWG; \citealt{albrecht_et_al_2009}), defining our Fisher matrix as:
\begin{equation}
\label{eq:fish}
{\cal F}_{ij} = \sum_b \frac{1}{\sigma_b^2}\frac{\partial f_b}{\partial p^i}\frac{\partial f_b}{\partial p^j},
\end{equation}
where $f$ is some observable measured at some $b$ values, $\sigma^2$ is the variance in a 
measured value of $f$, $p^i$ are cosmological parameters, and
we sum over all measured $f_b$ values.  

We propagate our power spectrum measurements into constraints on the Hubble parameter $H(z)$
and angular diameter distance $D_A$
using this formalism, where $f = \Delta^2(k)$, our measured power spectrum, $b = k$, the set
of $k$-modes we measured, and $p^i = [H(z),D_A]$.
The derivatives with respect to $D_A$ and $H(z)$ are straightforward to calculate, as they affect
our measurements through the $X$ and $Y$ parameters defined in equations \ref{eq:X} and \ref{eq:Y}.
In effect, changing $H(z)$ or $D_A$ changes the $k$-modes sampled by BAOBAB.
For our calculations, we exclude $k$-modes deemed contaminated by our foreground
model of \S\ref{sec:dspec}.  We also model the nonlinear degeneration of higher $k$-modes
using the elliptical Gaussian formula from \citet{seo_and_eisenstein_2007}.
To isolate the constraints provided by the BAO features from the broad-band shape of the
power spectrum, we again remove a model fit using the transfer function from
\citet{eisenstein_and_hu_1998}.  We split our data into three redshift bins centered at 
$z = 0.67,0.89$ and 1.18.  Although our frequency coverage is continuous between 
$z = 0.58$ and 1.37, we find that there is minimal penalty for using only three bins in a Fisher matrix
study.  The result of this calculation is that our fiducial 3-declination, 180-day 
integration yields measurements of 
$H(z)$ with an error ranging from 9\% to 4.5\% across our three redshift bins, from low to high redshift, and measurements of $D_A$ with errors effectively constant at 17\% over the same range
(note that errors $\gtrsim 5$\% should
be understood in the usual formal Fisher matrix sense -- these measurements
would not correspond to significant detections, taken in isolation).  
The particularly poor constraints on $D_A$ come from the loss of modes due to foreground emission;
we further explore the effects of our foreground model in \S\ref{sec:shortcomings}.
Taking
correlations between $H(z)$ and $D_A$ into account, these measurements amount to $2.5\%$ to $4.5\%$
errors on a ``dilation factor," which scales $D_A$ and $H^{-1}(z)$ in proportion.  Expressed
as a single constraint on the $z \sim 1$ distance scale, these measurements correspond to a
dilation factor error of $1.8\%$.
The exact uncertainties are given in Table \ref{tab:constraints}.

\begin{table*}[ht]\centering
\caption{Percent errors on the distance scale from BAO measurements,  
for a three declination BAOBAB survey.
The correlation is the correlation coefficient between the $H(z)$ and $D_A$ measurements.  
$R$ is the ``dilation factor," a single estimate of
the distance scale which scales $D_A$ and $H^{-1}(z)$ in proportion.
BOSS-LRG constraints come from
\citet{schlegel_et_al_2009} and BOSS-Ly$\alpha$ Forest constraints come from the method of
\citet{mcdonald_and_eisenstein_2007} and \citet{dawson_et_al_2012}.}
\begin{tabular}{l|ccccc}
Survey & Redshift & $H(z)$ Error & $D_A$ Error & Correlation & $R$ Error\\
\hline\hline
BAOBAB & 0.67 & 8.9\% & 17.1\% & 0.71 & 4.4\% \\
BAOBAB & 0.89 & 6.1\% & 16.4\% & 0.72 & 3.3\% \\
BAOBAB & 1.18 & 4.5\% & 17.5\% & 0.73 & 2.6\% \\
\hline
BOSS-LRG & 0.35 & 1.8\% & 1.0\% & 0.41 & 0.7\% \\
BOSS-LRG & 0.6 & 1.7\% & 1.0\% & 0.41 & 0.7\% \\
\hline
BOSS-Ly$\alpha$ & 2.5 & 3.1\% & 7.4\% & 0.58 & 2.0\% \\
\end{tabular}
\label{tab:constraints}
\end{table*}

It is also straightforward to propagate these errors on $H(z)$ and $D_A$ into errors on underlying
cosmological parameters through the Fisher matrix formalism of equation \ref{eq:fish}.
In this case, $f$ is now $H(z)$ or $D_A$, and $b$ corresponds to the redshift bin it was measured in.
$H(z)$ is given by:
\begin{align}
H^2(z) = &\ H_0^2\Big[ \Omega_m(1+z)^3 + \Omega_k(1+z)^2 + \nonumber \\
& \Omega_{DE}\ \rm{exp}\left(3\int_0^z\frac{dz'}{1+z'}[1+w(z')]\right)\Big].
\end{align}
$D_A$ is given by:
\begin{equation}
D_A(z) = \frac{1}{1+z}\int_0^z\frac{cdz}{H(z)}
\end{equation}
The parameters of interest, $p^i$, are now the underlying cosmological parameters.  We use
the parameterization of the JDEM FoMSWG, 
which include $w_m, w_b, w_k$ and $w_{DE}$ 
($w_X = \Omega_Xh^2$, where $m, b, k$, and $DE$ correspond to the matter, baryon, curvature and
present day dark energy density, respectively) although we use the simpler 2-component form
for the dark energy equation of state:
\begin{equation}
w(a) = w_0 + (1-a)w_a.
\end{equation}
Following the convention of Dark Energy Task Force report 
\citep{albrecht_et_al_2006}, we marginalize over all other parameters after combining 
our Fisher matrices with constraints from other experiments, to create $2\times2$ matrix representing
constraints on $w_0$ and $w_a$.  
As a Figure of Merit we use FoM = $|{\cal F'}|^{1/2}$, where $\cal F'$
is our original Fisher matrix $\cal F$, marginalized to a 2D $(w_0,w_a)$-space; this FoM
is proportional to the inverse of the error ellipse area in the $w_0-w_a$ plane.

\begin{figure}\centering
\includegraphics[width=3.5in]{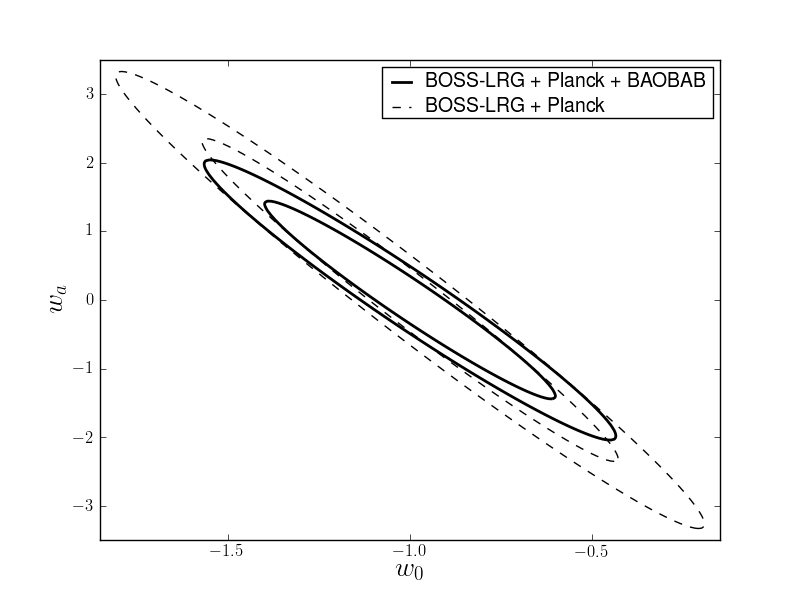}
\caption{1 and $2\sigma$ error ellipses in the $w_0-w_a$ plane for various 
surveys.  The dotted line shows constraints from Planck, BOSS-LRG 
and a 5\% error on $H_0$; the solid line shows the effect of including a 
1440-hour integration on three independent declination fields with a 132-tile 
BAOBAB array.  These error ellipses correspond to Figures of Merit of 8.7 and 
16.6, respectively.}
\label{fig:fish}
\end{figure}

As a baseline for current dark energy constraints,
we use the JDEM FoMSWG predictions for the Planck satellite, combined with constraints from the 
BOSS-LRG survey listed in Table \ref{tab:constraints} and a 5\% error on $H_0$.  
This combination of experiments yields an FoM
of 8.7.  
Including the measurements from our fiducial BAOBAB observation increases this
FoM value to 16.6.  
For comparison, BAOBAB combined with Planck and the $H_0$ constraint only yields a FoM of 4.4; 
the strength of BAOBAB therefore lies in adding complementary high-redshift information to the
BOSS-LRG survey: 
the high-redshift constraints from BAOBAB can significantly improve our current
measurements of the dark energy equation of state.  
(And, since the BOSS experiment is already underway and yielding high-quality data
\citep{dawson_et_al_2012}, this is the more interesting comparison to make.)
These constraints are plotted as 1 and 2$\sigma$ error ellipses in Figure
\ref{fig:fish}.  
If we include the Ly$\alpha$ forest
survey of BOSS, our baseline constraint FoM becomes 17.8 which is improved to 23.0 with the inclusion
of BAOBAB data.  
Even when the BOSS-Ly$\alpha$ forest constraints between redshifts 2 and 3 are 
added, BAOBAB still provides valuable information, serving to increase the FoM by $\sim 25\%$.

\section{Discussion}
\label{sec:discussion}

We break this discussion into two parts.  In \S\ref{sec:shortcomings}, we consider two 
components of our analysis which may be overly simplistic: our model of the BAOBAB primary
beam as a frequency-independent Gaussian, and our foreground emission model.  We discuss the effects 
any shortcomings in these models could have on our conclusions. In \S\ref{sec:improving},
we consider the fact that sample variance is the dominant source of uncertainty in our
measurements, and present future approaches that could improve the dark energy constraints possible 
with the BAOBAB instrument.

\subsection{Potential Shortcomings in the Analysis}
\label{sec:shortcomings}
A effect that could modify the predictions presented in this
work is the use of an overly simplistic model for the BAOBAB beam: a frequency independent
Gaussian.  Although this model is sufficient to calculate sensitivities, the effects of
a more realistic beam model on the delay spectrum foreground removal technique will necessitate
further investigations.  The principle cause for concern is from the existence of frequency-dependent
grating-lobe structure associated with the tile.  These sidelobes can introduce apparent 
frequency structure into otherwise spectrally
smooth foreground emission.  If this structure represents a significant increase in the size
of the delay-space convolving kernel, it will move the foreground contaminated region to higher $k$ values.  If the size of the effect is
large enough to push foreground emission on the shortest baselines beyond the first BAO peak,
the predicted cosmological constraints could be reduced.

Note, however, that the frequency-independent Gaussian beam is not as bad an assumption as it
might first appear.  Our choice to neglect the frequency evolution of the beam is partially
motivated by experience with the PAPER dipole beam, which, like BAOBAB, uses a modified 
dual-polarization ``sleeved" dipole design to limit the frequency evolution of the beam to
only $\sim10\%$ over the $120-180$~MHz band \citep{pober_et_al_2012}.
Of more concern are the grating lobes introduced by beamforming in tiles.
If we were trying to image the sky, ignoring the grating lobes
would be unjustified.  With the delay spectrum approach, though, the issue is not
the existence of the sidelobes, but their frequency dependence.  If the frequency Fourier transform
of the beam pattern is particularly broad --- corresponding to rapid evolution of the beam pattern
with frequency --- then foreground emission will have a similarly broad footprint in delay space,
compromising the 21cm signal.  Of course, the grating lobes themselves will change position as a 
function of frequency, introducing additional structure not in the PAPER beam.  
However, as argued in P12b, it is difficult for an element
only several wavelengths across to possess such frequency structure.  As stated in \S\ref{sec:dspec},
we will postpone a detailed investigation of these effects to a future paper with empirical
studies of the beam shape, and focus here
on array sensitivities, for which the Gaussian model is sufficient.

\begin{figure}\centering
\includegraphics[width=3.5in]{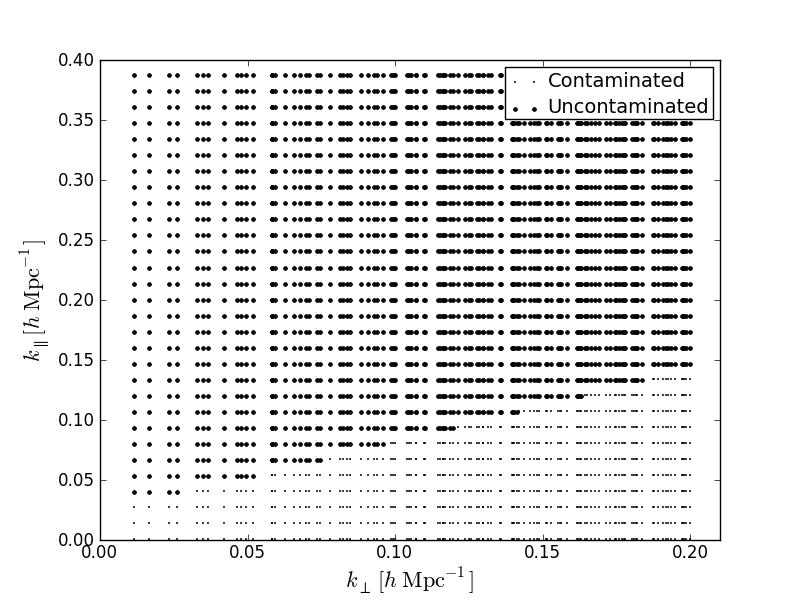}
\caption{The sampling pattern of BAOBAB-132 at $z = 0.89$ in the 
($k_{\perp}$-$k_{\parallel}$)-plane.  Modes are 
marked as either uncontaminated or 
contaminated by foregrounds using our model of \S\ref{sec:dspec}.
Foregrounds limit $\mu$ to, e.g., $\gtrsim0.65$ for 
$k\sim 0.1$, leading to the relatively poor $D_A$ measurement and high correlation.
} \label{fig:fgloss}
\end{figure}

As noted in \S\ref{sec:baobab128}, our constraints on $D_A$ are significantly limited by foreground
emission.  We illustrate this effect in Figure \ref{fig:fgloss}, which shows the sampling pattern
of BAOBAB-132 at 750 MHz in the ($k_{\perp}$-$k_{\parallel}$)-plane, highlighting those modes 
discarded as contaminated by our foreground model of \S\ref{sec:dspec}.  
Foreground
contamination effectively excludes modes where $k_{\perp} \lesssim k_{\parallel}$, i.e., transverse
modes.  This has the effect of significantly degrading the achievable constraints on $D_A$.

While our foreground model presented in \S\ref{sec:dspec} is empirically motivated, 
accurate predictions for foreground emission
will need to come from early BAOBAB arrays or other 21cm experiments.  We briefly explore the effect
of changing the size of the additive component of our foreground model arising from the
spectral smoothness (or lack thereof) the emission.  For our fiducial model, this term has a
magnitude of $k_{\parallel} = 0.02~h{\rm Mpc}^{-1}$; as test cases, we analyze the constraints
obtainable with BAOBAB if this term is changed by $\Delta k_{\parallel} = 
\pm~0.01~h{\rm Mpc}^{-1}$.  
This has the effect of moving the cutoff between contaminated and uncontaminated modes in Figure
\ref{fig:fgloss} up or down by $0.01~h{\rm Mpc}^{-1}$.

The effect of increasing this term (i.e., if foreground emission is not as spectrally smooth as 
predicted) is to degrade the significance of our BAO detections in each of the redshift bins by $\sim 0.4\sigma$, our $H(z)$,
$D_A$, and $R$ constraints by $\sim0.5$, $\sim5$, and $\sim0.3$ percentage points, respectively.
Reducing the foreground emission footprint in $k$-space (i.e., if foregrounds are spectrally
smoother than predicted) has similar effects with the opposite sign: the significance of
our BAO detections in each band are increased by $\sim 0.3\sigma$, our $H(z)$, $D_A$, and $R$ constraints are improved
by $\sim0.5$, $\sim5$, and $\sim0.3$ percentage points, respectively.  From this analysis, it is clear
that foreground emission can significantly alter the achievable constraints on $D_A$, but 
ultimately, the success of BAOBAB will not be determined by the details of foreground spectral
properties.

\subsection{Improving Our Constraints}
\label{sec:improving}

As discussed in \S\ref{sec:baobab128}, sample variance dominates the
uncertainties in power spectrum measurements from BAOBAB.  However, measuring
a new set of independent modes is not a trivial undertaking.  By using a close-packed array,
BAOBAB completely samples the $uv$-plane out to some maximum baseline length.  Therefore a simple
array rotation or reconfiguration will not yield new samples.  Furthermore, in our highest redshift bin of $z = 1.18$, the longest baseline in the 132-tile array probes a transverse $k$-mode of 
$k_{\perp} = 0.15~h{\rm Mpc}^{-1}$ (and at lower redshift, this longest baseline corresponds
to an even larger value of $k_{\perp}$).  At this scale and smaller, most BAO information is being 
lost to non-linear damping effects.  Therefore, while a larger array will beat down thermal noise
faster, the cosmological returns from increasing the array size beyond 
$\sim128$ tiles are limited, since effectively no new modes with significant BAO information will be
probed.

As an upper-limit to the constraints obtainable with a single-declination, $\sim1200$ square degree, 
132-tile BAOBAB observation (as opposed to our fiducial observation targeting three declination 
fields)
we compute the results of a completely sample variance limited survey, i.e., one where thermal noise
uncertainties have been set to 0 (although modes are still excluded using our foreground
emission model).   
This sample variance
limited 132-tile BAOBAB observation yields distance scale uncertainty of $\sim1.5\%$, averaged over our whole band (compared with $1.8\%$ for our
fiducial 1-year observation of 3 independent declination fields).  
For comparison, a 10-year (14,400 hour), one declination observation with the same
array yields a distance scale uncertainty of $1.9\%$.
Therefore, while better constraints can come from a longer observation, obtaining measurements of 
new modes to beat down sample variance is clearly the optimal way to proceed.  There are two ways
forward to achieve this goal: map a different volume of the universe (as with our fiducial 
experiment) or recover foreground corrupted modes.  We consider each of these approaches in turn.

Since BAOBAB is a zenith-pointing, drift-scanning telescope, to map a new area of the sky, it will
either need to be relocated to a different latitude or tilted to point towards a different patch of sky.  Either option
is potentially feasible, as even a 132-tile BAOBAB experiment spans less than 20 meters.  With
a primary beam full-width half-max of $\sim 15^{\circ}$, there are $\sim10$ 
independent pointings
in declination that BAOBAB can target.  
Since our fiducial observation already targets three
declination fields, mapping every declination could in principle yield up to 
$\sim70\%$ reductions
in the error bars over the results presented. 

The other way to potentially measure new modes and beat down sample variance is to recover samples
we have considered corrupted by foregrounds.  There are two ways foregrounds 
compromise BAOBAB
observations.  The first is the limited observing time per-day, set by Galactic emission,
which we have treated as irreparably corrupting all samples, even those in principle
recoverable with the delay transform.  If it is possible to observe all 24 hours of right ascension,
as opposed to the 8 considered here, the constraints from a single observing season will increase
by a factor of $\sqrt3$.  While it is unlikely that all 24 hours of right ascension
will be workable, our fiducial value of 8 hours per day, motivated by observations
with present EoR experiments, may well
be conservative, since Galactic synchrotron emission has significantly fallen in brightness
compared to EoR frequencies.

Even when observing a ``cold patch," foregrounds corrupt large scale $k$-modes with a 
footprint moving to smaller scales as baseline length increases (\S\ref{sec:dspec}).  If these
modes could be retrieved, they could significantly increase the volume of Fourier space
that BAOBAB can probe.  As an upper-limit, we calculate the obtainable power spectrum constraints 
ignoring all foreground contamination.  The result is that a 1-year (1440-hour) observation
in each of three independent declination fields
yields a distance scale uncertainty of 1.4\% combined over the entire redshift range, an increase of $\sim25\%$
over the same observation including foreground emission.  
In particular, we note that a foreground-free observation yields
errors of $\sim5\%$ on $D_A$ at redshifts of 0.67, 0.89 and 1.18, respectively --- 
a factor of $\gtrsim3$ improvement over the predictions for an observation including the effects of foreground emission.
While
an analysis of foreground removal techniques is beyond the scope of this work, 
this result is
suggests that foreground removal may be the way to improve constraints on $D_A$.

As an order of magnitude estimate, we can  consider whether a foreground 
removal or
subtraction scheme might be more effective in the BAO band than at EoR frequencies.  At 
$k\sim0.1~h\rm{Mpc}^{-1}$, the 750-MHz BAO 21cm power spectrum reaches $\sim~3\times10^{-3}~\rm{mK}^2$,
compared with a peak brightness at EoR frequencies of 150 MHz reaching $\sim~10~\rm{mK}^2$.  
The steep spectrum Galactic synchrotron emission has a spectral index of -2.5, and so will fall
by a factor of $(5^{-2.5})^2 = 3.2\times10^{-4}$ in units of temperature squared.  (Extragalactic point sources are
less steep spectrum, and so will not fall off in brightness as steeply.  Therefore, this estimate
can be considered a lower limit on the foreground-to-signal ratio).  
Roughly speaking, then, the foreground-to-signal ratio is unchanged compared with EoR experiments,
suggesting
that a foreground isolation scheme like the delay-spectrum technique is still likely
the most viable approach for first-generation experiments limited in collecting area.

\section{Conclusions}
\label{sec:conclusions}

In this work we have presented a concept for a new experiment using the redshifted 21cm 
line of neutral hydrogen to probe cosmology at $z\sim1$.  The BAO Broadband and Broad-beam
Array (BAOBAB) will incorporate both the hardware and analysis infrastructure developed for
21cm experiments at higher redshifts.

The hardware design will borrow heavily from the Precision Array for Probing the Epoch
of Reionization (PAPER) and the Murchsion Widefield Array (MWA), 
using a scaled version of the PAPER dipoles as a feed element, tiling dipoles as done by the MWA,
and modifying the CASPER FPGA/GPU PAPER correlator to perform full
dual-polarization cross-correlations of all elements.  Significant improvements to the
system temperature will be brought about through state-of-the-art uncooled, 
low-noise amplifiers.  Relative to PAPER, the collecting area will be substantially
increased through the use of tiles of 4 dipoles combined through a beamformer, as demonstrated
by MWA efforts.  Although we have largely avoided specific cost-estimates, it is fair to say that
this infrastructure is obtainable at a fraction of the cost of the $\sim100$-million dollar 
ground-based optical redshift surveys.

On the analysis side, BAOBAB will use the maximum redundancy
configurations and delay spectrum foreground removal techniques presented in 
\citet{parsons_et_al_2012a} and \citet{parsons_et_al_2012b} to enhance sensitivity to
Fourier modes along the line-of-sight.  Motivated by the science of Baryon Acoustic Oscillations,
BAOBAB will utilize extremely close-packed arrays to maximize the number of short baselines.
The sensitivity calculations presented here
show that BAOBAB will achieve several milestone measurements over our anticipated staged deployment
process.  A $\sim32$-element BAOBAB system will yield high significance detections of the HI power
spectrum, and constrain the evolution of the cosmic neutral hydrogen fraction from $z = 0.5$ to
1.5 as well as the bias of DLAs.  Over the same wide redshift range, a $\sim128$-element system will allow for a first detection of the BAO features in the
power spectrum, and yield errors on the distance scale $R$ at the several percent level.  When combined
with our current constraints on dark energy, including those forthcoming from the BOSS and Planck 
experiments,
BAOBAB's measurements result in substantial increase in the Dark Energy Task Force Figure of Merit,
representing constraints on the nature and time evolution of dark energy over a wide range of cosmic history.

\acknowledgments{We thank our anonymous reviewer for helpful comments which have resulted in a significantly improved and clearer manuscript.}

\end{document}